\documentclass[fleqn,usenatbib]{mnras}
\usepackage{newtxtext,newtxmath}

\usepackage[T1]{fontenc}

\DeclareRobustCommand{\VAN}[3]{#2}
\let\VANthebibliography\thebibliography
\def\thebibliography{\DeclareRobustCommand{\VAN}[3]{##3}\VANthebibliography}

\usepackage{graphicx}	
\usepackage{amsmath}	
\usepackage[labelfont=bf]{caption}
\usepackage{adjustbox}
\usepackage[super]{nth}
\usepackage[]{siunitx}
\usepackage{subfig}
\usepackage{mathtools}
\usepackage{bm}
\usepackage{enumerate}
\usepackage{nth} 
\usepackage{xspace} 

\newcommand{\sn}{{\rm S/N}\xspace}

\newcommand{\lya}{Ly$\alpha$\xspace}

\newcommand{\pq}{\ensuremath{P_\mathrm{q}\xspace}}

\newcommand{\dsq}{\ensuremath{{\rm deg}^2 }}
\newcommand{\chisq}{\ensuremath{\chi^2}\xspace}

\newcommand{\mcs}{\texttt{MergedClassStat}\xspace}
\newcommand{\given}{\,\middle\vert\,}
\newcommand{\ks}{\ensuremath{K_{\rm s}}\xspace}
\newcommand{\zy}{\ensuremath{Z-Y}\xspace}
\newcommand{\yj}{\ensuremath{Y-J}\xspace}

\newcommand{\hk}{\ensuremath{H-\ks\xspace}}
\newcommand{\yk}{\ensuremath{Y-\ks}\xspace}
\newcommand{\myrange}[2]{{#1}\,\textrm{--}\,{#2}\xspace}
\newcommand*{\diff}{\ensuremath{{\rm d}}}

\title[VIKING $z\gtrsim6.5$ quasars]{A complete search for redshift \mbox{\boldmath{$z \gtrsim 6.5$}} quasars in the VIKING survey}

\author[R. Barnett et al.]{
R. Barnett$^{1}$,
S. J. Warren$^{1}$\thanks{E-mail: s.j.warren@imperial.ac.uk},
N. J. G. Cross$^{2}$,
D. J. Mortlock$^{1,3,4}$, 
X. Fan$^{5}$,
F. Wang$^{6}$
and P. C. Hewett$^{7}$
\\
$^{1}$Astrophysics Group, Blackett Laboratory, Imperial College London, London SW7 2AZ, United Kingdom\\
$^{2}$Wide-Field Astronomy Unit, Institute for Astronomy, School of Physics and Astronomy, \\
University of Edinburgh, Royal Observatory, Blackford Hill, Edinburgh EH9 3HJ, UK\\
$^{3}$Department of Mathematics, Imperial College London, London SW7 2AZ, UK\\
$^{4}$Department of Astronomy, Stockholm University, Albanova, SE-10691 Stockholm, Sweden\\
$^{5}$Steward Observatory, University of Arizona, 933 N. Cherry Avenue, Tucson, AZ 85721, USA\\
$^{6}$Department of Physics, Broida Hall, University of California, Santa Barbara, CA 93106, USA\\
$^{7}$Institute of Astronomy, University of Cambridge Madingley Road, Cambridge CB3 0HA, UK\\
}

\date{Accepted XXX. Received YYY; in original form 2020 August 26}

\pubyear{2020}

\begin{document}
\label{firstpage}
\pagerange{\pageref{firstpage}--\pageref{lastpage}}
\maketitle

\begin{abstract}
We present the results of a new, deeper, and complete search for
high-redshift $6.5<z<9.3$ quasars over 977\,deg$^2$ of the VISTA
Kilo-Degree Infrared Galaxy (VIKING) survey. This exploits a
new list-driven dataset providing photometry in all bands $ZYJH\ks$,
for all sources detected by VIKING in $J$. We use the Bayesian model
comparison (BMC) selection method of Mortlock et al., producing a
ranked list of just 21 candidates. The sources ranked 1, 2, 3 and 5
are the four known $z>6.5$ quasars in this field. Additional
observations of the other 17 candidates, primarily DESI Legacy Survey
photometry and ESO FORS2 spectroscopy, confirm that none is a
quasar. This is the first complete sample from the VIKING survey, and
we provide the computed selection function. We include a detailed
comparison of the BMC method against two other selection methods:
colour cuts and minimum-\chisq SED fitting. We find that: i) BMC
produces eight times fewer false positives than colour cuts, while
also reaching 0.3 mag. deeper, ii) the minimum-\chisq SED fitting
method is extremely efficient but reaches 0.7 mag. less deep than the
BMC method, and selects only one of the four known quasars. We show
that BMC candidates, rejected because their photometric SEDs have high
$\chi^2$ values, include bright examples of galaxies with very strong
[OIII]$\lambda\lambda$4959,5007 emission in the $Y$ band, identified
in fainter surveys by Matsuoka et al. This is a potential contaminant
population in {\em Euclid} searches for faint $z>7$ quasars, not
previously accounted for, and that requires better characterisation.
\end{abstract}




\section{Introduction}
\label{sec:intro}

Quasars at redshift $z>6.5$ are useful probes of early supermassive black hole growth and the epoch of reionisation.
Since the discovery of the first such quasar \citep{Mortlock2011},
the current tally of $z>6.5$ quasars stands at almost fifty,
with discoveries made using a wide range of near-infrared (NIR) surveys
\citep{Venemans2013,Venemans2015a,Matsuoka2016,Matsuoka2018a,Matsuoka2018b,Matsuoka2019,
	Decarli2017,Koptelova2017,Tang2017,Reed2017,Reed2019,Wang2017,Wang2019,Songaila2018,Pons2019,Fan2019,Yang2019,Yang2020}.

Discovering $z>6.5$ quasars remains challenging, not least due to their space density, which declines strongly with increasing redshift.
The decline of the number density of quasars brighter than a specific absolute magnitude $M_{1450}$ is often parametrised as
\begin{equation}
\rho\left(z,<M_{1450}\right)=\rho\left(z_0,<M_{1450}\right)\,10^{k\left(z-z_0\right)},
\end{equation}
where $z_0$ is an arbitrary reference redshift. A comprehensive measurement of the QLF at $z\sim6$ was made by \citet{Jiang2016},
who used a complete sample of 47  SDSS quasars, $5.7 < z < 6.4$,
measuring a rapid fall in quasar number density over $z=\myrange{5}{6}$, 
with $k = -0.72\pm 0.11$. For $M_{1450}=-26$, they measure $\rho\sim 1$\,Gpc$^{-3}$ (comoving) at $z=6$,
corresponding to a surface density of approximately one object per 100\,\dsq.

In addition to the low numbers of quasars, selection of $z>6.5$ quasars is severely hampered 
by contamination from intervening populations: cool stars and brown dwarfs (henceforth MLTs); 
and compact early-type galaxies (henceforth ETGs) at intermediate redshifts ($z\sim\myrange{1}{2}$), misclassified at low S/N as of stellar morphology. These populations are far more abundant than,
and have similar NIR colours to, the target quasars \citep[e.g.,][]{Hewett2006}\footnote{
	A further known class of contaminant which we do not explicitly treat in this work is extreme FeLoBALs \citep[e.g.,][]{Hall2002}, where strong MgII absorption can produce a sharp continuum break. 
	These are considerably less common than MLTs and ETGs, but more common than $z>6.5$ quasars.}. Consequently, colour-selected samples of candidates 
are dominated by contaminants,  especially as quasar searches move to lower \sn to maximise the number of discoveries. Because of this many searches have stopped after identifying the more obvious bright quasars, and have not persisted to the point of following up all candidates to produce a complete sample with a computed selection function, i.e., the measured completeness as a function of absolute magnitude and redshift. The measurement by \citet{Wang2019} of the space density at $z>6.5$, using the DESI Legacy Survey \citep{Dey2019} and Pan-STARRS \citep{Chambers2016}, is the only such analysis at these redshifts based on a complete sample. They measured $k=-0.78\pm0.18$ between $z=6$ and $z=6.7$, consistent with the rate of decline measured over $5<z<6$.

The main purpose of the current paper is to use the VISTA Kilo-Degree Infrared Galaxy (VIKING) survey \citep{Edge2013} to produce a complete sample of $z>6.5$ quasars (henceforth HZQs; high-redshift quasars), with a computed selection function, that reaches fainter absolute magnitudes and higher redshifts than the survey of \citet{Wang2019}.
Improving and extending measurements of the quasar luminosity function (QLF) to fainter luminosities and higher redshifts is an important aim,
as such studies will constrain models of the formation 
and growth of supermassive black holes at early times \citep[e.g.,][]{Willott2010,Jiang2016},
and are also important for designing future surveys for quasars at $z\sim7$ and beyond,
such as for \textit{Euclid} \citep{Barnett2019}.
The VIKING survey has already been searched using colour cuts, 
yielding four HZQs \citep{Venemans2013, Wang2017}.
To improve upon these previous searches
we exploit a new list-driven photometric catalogue, 
which provides aperture-corrected aperture photometry in all five available bands $ZYJH\ks$ 
for every VIKING source detected in the $J$ band \citep{Cross2014}. As explained in the next section this database has advantages for searches for HZQs.

To search the VIKING database we use an updated version of the Bayesian model comparison (BMC) technique developed by \citet{Mortlock2012}. This method has been adapted for Subaru data by \citet{Matsuoka2016} and used in their SHELLQs surveys. The same method has been adapted by \citet{Pipien2018} and applied to CFHT data.
In a previous paper \citep{Barnett2019} we compared two different search methods, namely BMC and minimum-$\chisq$ SED fitting 
(e.g., \citealt{Reed2017}, applied to DES data). Using simulated datasets we showed that the BMC method is the more useful method as it reaches much deeper than the SED fitting method, while still being highly efficient. A third search method is the use of simple colour cuts, as applied by \citet{Venemans2013} in their search of the VIKING survey. Although we do not use the candidate lists, for the sake of comparison of the efficiency and depth of the methods we also produce candidate lists using the SED fitting and colour cuts methods. This extends the comparison of methods undertaken by \citet{Barnett2019}, and has the advantage of using real data.

The paper is structured as follows. In Sect.~\ref{sec:listdriven} we give an overview of the VIKING survey, 
and the new list-driven dataset. In Sect.~\ref{sec:selection} we describe the BMC selection method used for the survey. We also detail the two other methods, SED fitting and colour cuts, that are used in the comparison of techniques. We present the results from our BMC HZQ search in Sect.~\ref{sec:search}.
We compute the selection functions for the BMC search and for the other two methods in Sect.~\ref{sec:simulations}.
In Sect.~\ref{sec:mltell} we test that the models of the contaminating populations are reasonable by creating a synthetic survey and comparing, for the three selection methods, the numbers of simulated candidates to the numbers of candidates found in the real data.
We summarise in Sect.~\ref{sec:end}. 
All magnitudes and colours quoted are on the Vega system,
the default for the VISTA telescope,
unless otherwise stated. 
The AB conversions used in this work for $Z,Y,J,H,\ks$ are respectively
0.524, 0.618, 0.937, 1.384, 1.839\footnote{
These conversions are provided on the Cambridge Astronomical
Survey Unit (CASU) website (\url{http://casu.ast.cam.ac.uk/surveys-projects/vista/technical/filter-set}), and calculated following \citet{Hewett2006}.}.

Where required we have used cosmological parameters 
$h$ = 0.7, $\Omega_{\rm M} = 0.3$, and
$\Omega_{\Lambda}$ = 0.7.

\section{VIKING survey}
\label{sec:listdriven}

VIKING is a medium-deep NIR survey 
covering $\sim1300\,\dsq$ 
in five broadband filters 
-- $Z,Y,J,H~{\rm and}~\ks$ -- 
with the VISTA telescope. VIKING datasets are named by the release date.
The list-driven photometry database used here is based on the dataset 20160406, which was the latest release at the time this project started. This contains a total of 835 framesets, and the majority have coverage in all filters. We use all the framesets that have, as a mimimum, observations in the $Z, Y, J$ filters. This subset contains 782 framesets, and has a footprint covering a total area of $988.7\,\dsq$. This calculation accounts for overlaps between framesets, and the fact that detector 16 is not used. With later releases the area has expanded to 1244.2\,deg$^2$, but a list-driven dataset does not currently exist of the additional area.

The standard VIKING catalogues are formed by 
merging lists of objects detected in each filter. 
The subsequent list-driven photometry \citep{Cross2014}
is motivated by the fact that 
the majority of HZQs will be too faint in the $Z$ band to be detected. This is a consequence of absorption by neutral hydrogen along the line of sight \citep{Lynds1971}, that increases with redshift \citep{Becker2001}, so that by $z=6.5$ nearly all the flux blueward of Ly$\alpha$ has been absorbed \citep[e.g.,][]{Barnett2017}.  
Consequently, in the standard catalogues the $Z$ band will only provide a flux limit for HZQs. 
For such sources, 
a flux measurement provides much more information than an upper limit \citep{Mortlock2012}.
Our new list-driven catalogue is produced 
by performing aperture photometry 
in each band at the position of 
every object that is detected in the VIKING $J$ band, and applying an aperture correction appropriate for a point source. These are therefore total magnitudes. The aperture corrections are measured from bright stars in the same field.
Further details of the creation of the list-driven catalogue will be provided in a forthcoming paper (Cross et al., in prep.; see also \citealt{Ross2020}).

\begin{figure}
	\centering
	\includegraphics[width=9cm]{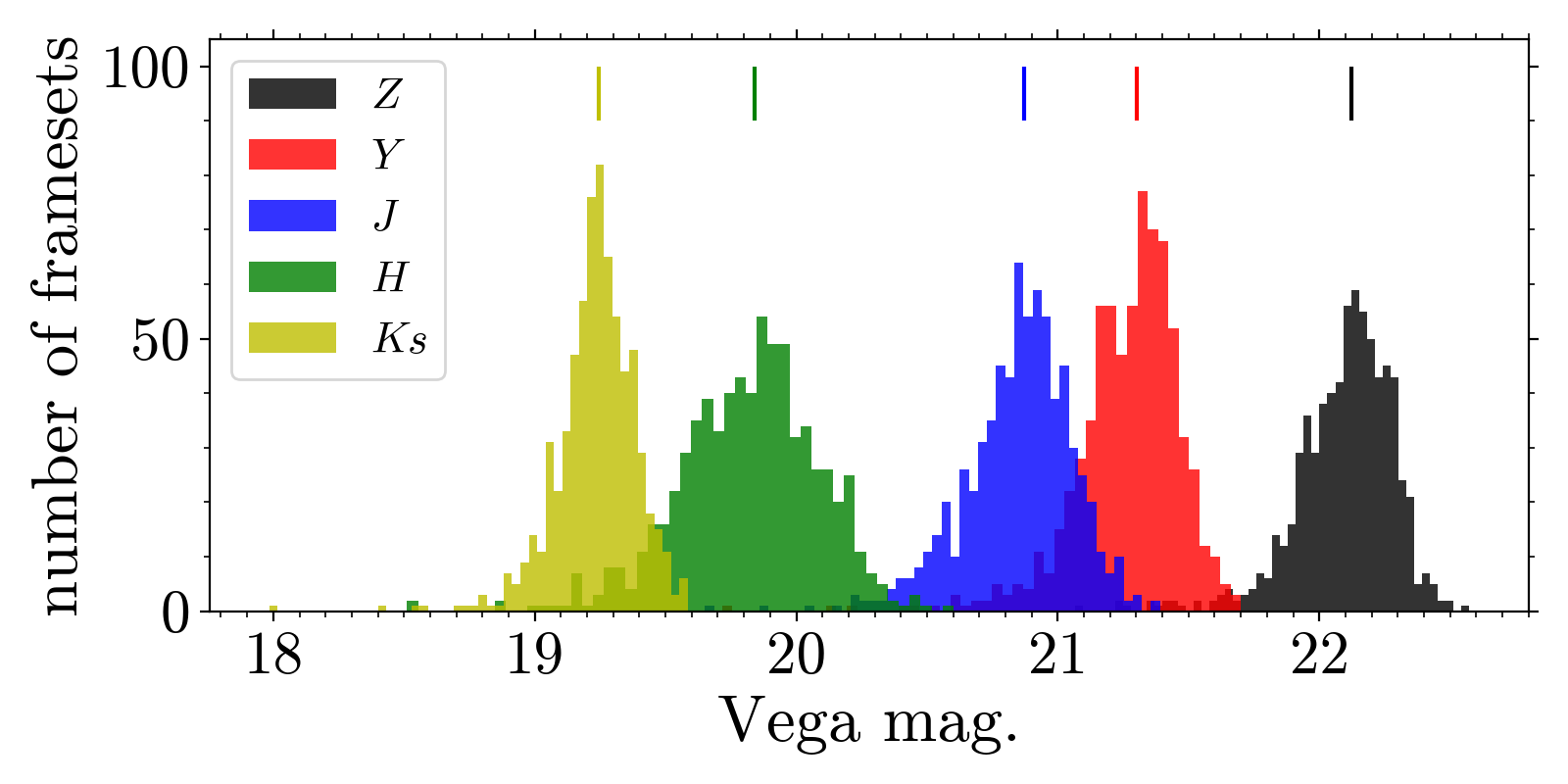}
	\caption{Distribution of $5\,\sigma$ frameset depths in each VIKING band. The dashes indicate the median value for each filter.}
	\label{fig:depths}
\end{figure}

The photometric depths vary across the survey. The distributions of depths in the different bands are plotted in Fig. \ref{fig:depths}. Here depth is quantified by the total magnitude of a point source that is detected at $5\sigma$ in a $\ang{;;2}$ diameter aperture. The median $5\,\sigma$ depths are $(Z,Y,J,H,\ks) = (22.1,20.3,20.9,19.8,19.2)$. These depths are some $~1.5\,$mag. deeper than the UKIDSS data used by \citet{Wang2019}.

Regions close to bright stars are excised from the dataset as we found that the VIKING photometry of sources is unreliable in these locations. We observe an excess of candidates near bright stars, which are therefore clearly false positives. The size of the region affected increases with the brightness of the star. We used the 2MASS catalogue \citep{Skrutskie2006} to quantify this, as the bright stars are saturated in VIKING. We drilled holes around stars brighter than $J=11,$ with a radius $R$ dependent on the 2MASS magnitude according to
$R = \ang{;;20}(11 - J_\mathrm{2MASS})$. 
There are $4.3\times10^4$ bright stars with $J_\mathrm{2MASS}<11$ in the VIKING footprint and we remove 12.2\,deg$^2$ from the survey, leaving an effective area of 976.5\,\dsq.

\section{High redshift quasar selection}
\label{sec:selection}

In this section we provide the details of our search for HZQs that uses the BMC method. In Sect.~\ref{sec:other} we also list the details of the SED fitting and colour cuts techniques that are used in our comparison of methods. The BMC and SED fitting methods both require model colours of the contaminating populations. The BMC method additionally requires models of the surface density, as a function of apparent magnitude, of the contaminating populations and of the quasars. The populations modelled are the same as we used in the $Euclid$ study \citep{Barnett2019}, but adapted specifically for VIKING data, taking into consideration the different filter transmission curves and image quality. The full details of the of the models of the populations (colours and surface densities of quasars and contaminants) are provided in the Appendix, Sect.~\ref{sec:pops}. Nevertheless, we begin this section with a very brief summary of the models, as this is needed as background for understanding the selection methods. 

\subsection{Overview of population models}

\begin{figure}
	\centering
	\includegraphics[width=6cm]{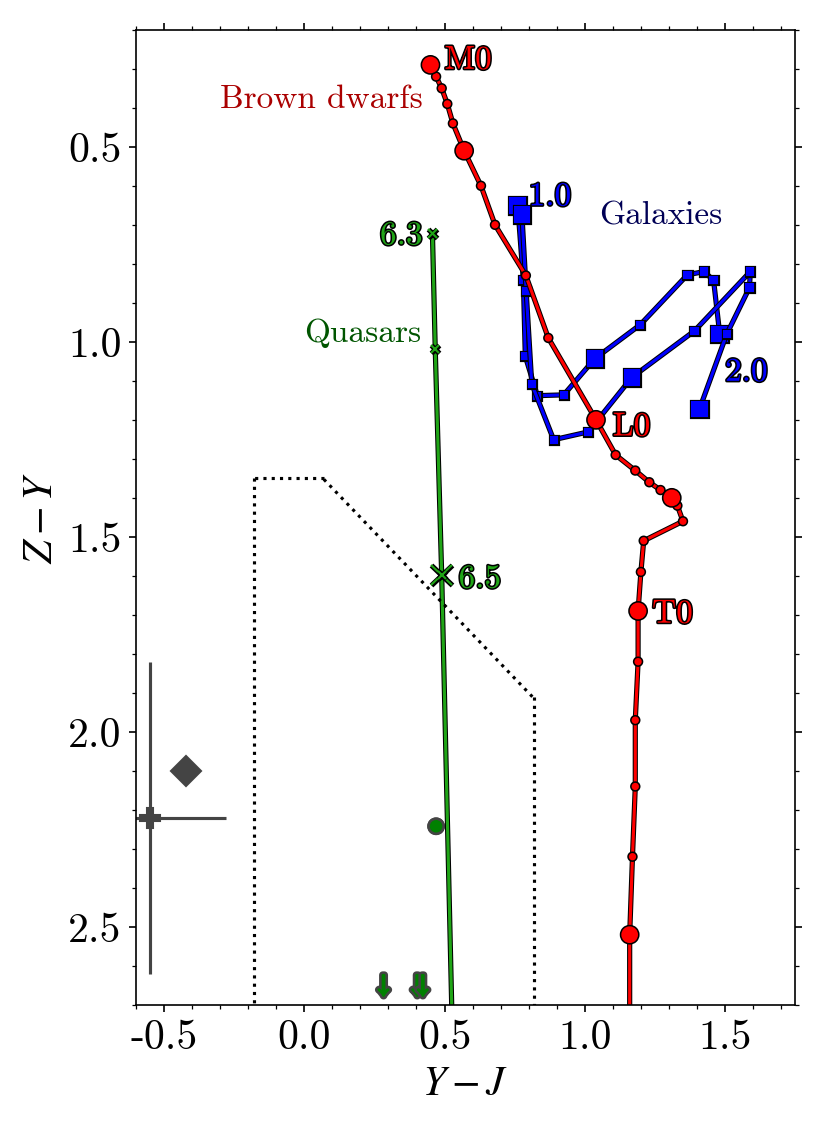}%
	\label{fig:ZYJ}%
	
	\includegraphics[width=6cm]{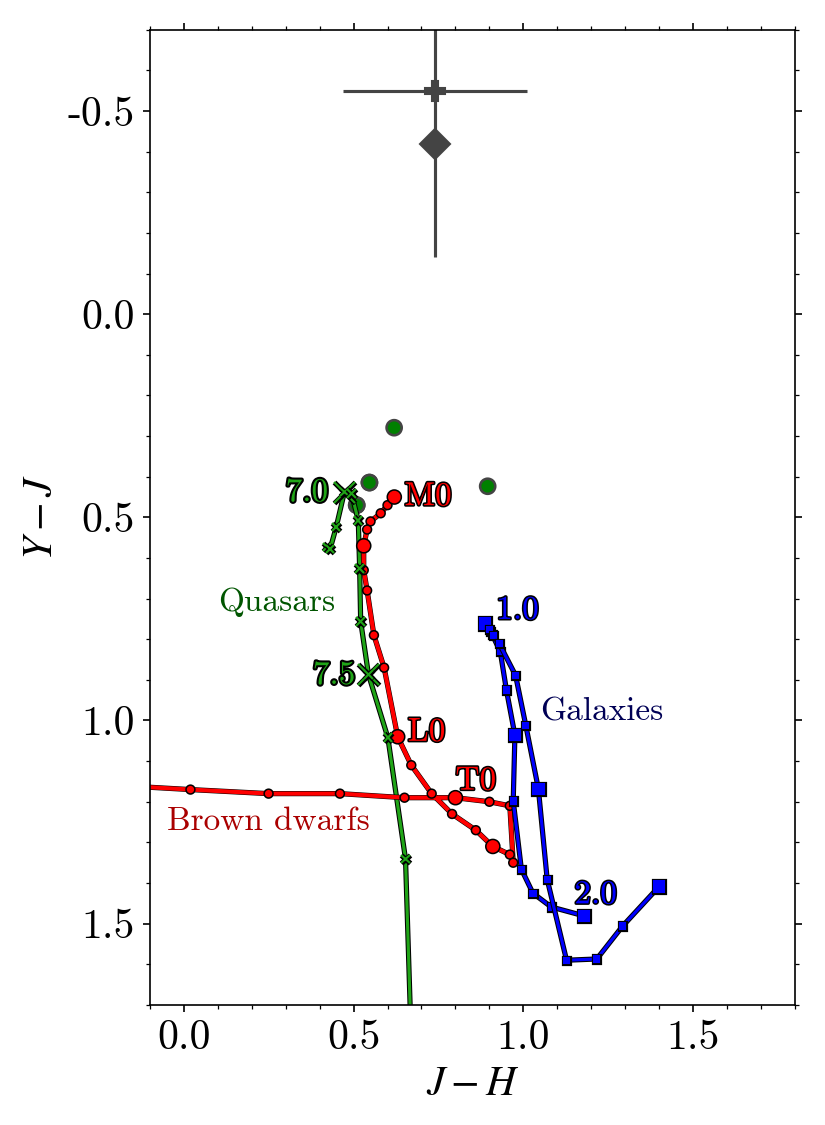}%
	\label{fig:YJH}%
	\caption{Colour-colour diagrams of relevant populations.
		{\em Green tracks (crosses):} HZQ model colours as a function of redshift, 
		spacing $\Delta z = 0.1$.
		{\em Red tracks (circles):} model MLT colours for each spectral type.
		{\em Blue tracks (squares):} ETGs
		($z_f=3$ and $z_f=10$), 
		spacing $\Delta z = 0.1$.
		Green circles are the known VIKING HZQs.
		The grey plus-sign indicates the emission-line galaxy VIK J1459--0321 (Sect.~\ref{sec:bmcsearch}).
		The grey diamond indicates the average colours of the high$-\chi^2$ rejected candidates, 
		(Sect.~\ref{sec:bmcsearch}).
		\textit{Upper:} $ZYJ$ colours. 
		Selection criteria of the colour-cuts method are shown as black dotted lines.
		Three of the known HZQs have $Z-Y>3$,
		and are indicated using arrows.
		\textit{Lower:} $YJH$ colours.
	}
	\label{fig:colourcolour}
\end{figure}

The colours of the three populations  (quasars, MLTs, and ETGs), are plotted in Fig.~\ref{fig:colourcolour}, showing the $Z-Y, Y-J$ and $J-H$ colours. The colours of the four known VIKING HZQs are also shown. We have used the convention of blue-top-left, red-bottom-right in these plots, as used in \citet{Sandage1965}, the paper that initiated multicolour searches for quasars. At high S/N the three populations are easily mutually distinguishable. At low S/N extreme outliers from the two contaminating populations, which outnumber quasars by orders of magnitude, can have the same observed colours as the target quasars.

The modelling of the colours of the quasar population utilises nine different spectral types, which are combinations of three different continuum slopes and three emission-line strengths. The nine different types are used in the SED-fitting selection method, and are also used (with appropriate weights) in computing the selection functions for all the methods. However the BMC selection method itself uses only a single quasar type (the typical SED). The significance of this choice is discussed when describing the BMC method (Sect.\ref{sec:bmc}). The modeled surface density of quasars uses the measured luminosity function at $z=6$, with a declining space density towards higher redshift.

The colours of MLT dwarfs, covering the spectral range M0 to T8, were determined from a combination of measured colours of sources classified by spectroscopy, as well as calculation of synthetic colours from spectra. We assume that the space density falls off exponentially with height from the Galactic mid-plane, and the adopted space density of each spectral sub-type comes from recent wide-field surveys for these populations.

ETGs at intermediate redshift $1<z<2$ have very red colours. The galaxy size is a strong function of luminosity so that fainter ETGs detected in VIKING are very compact and may be classified as stellar. Only very extreme outliers of this population can mimic the SEDs of HZQs, but the surface density of ETGs is more than four orders of magnitudes higher than that of HZQs. Combined with the difficulty of modelling the star/galaxy boundary, this means that this is the most difficult population to model accurately.

A relevant question is whether or not to allow for intrinsic spread in the colours of the different populations. For the MLTs, \citet{Skrzypek2015} showed that intrinsic spread was accurately accounted for by applying a dispersion of 0.05\,mag. in each band, added in quadrature to the photometric uncertainties. 
Therefore at very high S/N, brighter than any expected quasars, the colour tracks of the MLTs are narrower than they should be. However, should there be any unusually bright quasars in the survey this is not a problem, since of course they will be very well separated from the contaminating populations. In the expected magnitude range of quasars in the survey $J>18.5$ the photometric uncertainties dominate over the intrinsic spread. This means that the model for the MLTs adequately represents the population (adding 0.05 mag. in quadrature does not make a significant difference). A similar argument applies to the modelling of ETGs and quasars. For these there is some intrinsic spread in colour, of a discrete rather than a continuous nature. For the ETGs we split the population by adopting two different formation redshifts $z=3$ and 10. For the quasars we have the nine different spectral types. Again these individual colour tracks do not overlap at very bright magnitudes, but at $J>18.5$ all the sub-populations of respectively the quasars, and the ETGs, overlap because of the size of the photometric errors at these magnitudes. To confirm that additional intrinsic spread is not needed, we reran the selection functions for a case where additional intrinsic spread of 0.05\,mag. in each band was included, and found negligible difference. 

\subsection{Candidate selection using Bayesian model comparison} 
\label{sec:methods}

\subsubsection{Initial cuts}
To target HZQs, before applying the BMC algorithm we implement the following cuts:
\begin{enumerate}
	\item $\textrm{\sn}_J \geq 4$. 
	Given that our list-driven catalogue requires a $J$ band detection, 
	very few sources are removed by this step.
	We do not have an equivalent requirement in $Y$,
	which means our sensitivity in redshift extends to $z\sim9.3$.
	However, in a survey of the depth and area of VIKING the predicted yield
	is extremely small at such high redshift. 
	
	\item The particular field was observed in all three bands $ZYJ$.
	
	\item $\textrm{\sn} \geq 4$ in one additional band $Y,\,H\,{\rm or}\,\ks$. 
	In practice this ensures that a $J$-band detection corresponds to a real source.
	
	\item $\textrm{\sn}_Z < 4$, or $Z - Y \geq 1.5$.
	
	\item $-4 < \mcs\! < 2$. The parameter \mcs (MCS) is a morphology 
	statistic in VIKING, which we use to exclude identifiably extended
	sources. 
	
	\item Not on detector 16. Flat-fielding is not accurate for this CCD 
	due to a time-varying quantum efficiency and many detected sources are spurious. 
	A flag is automatically applied to VIKING sources from this detector \citep{Cross2012}. 
	The exclusion of detector 16 is accounted for in the quoted survey area
	and corresponds to a reduction in area of 6 per cent.

\end{enumerate}

The cut in $Z-Y$ colour was selected carefully to ensure that the selection function cuts off just below $z=6.5$. The redshift cut is sensitive to the emission line strength, since this affects the $Z-Y$ colour. This sensitivity to line strength is accounted for in computing the selection function, by the weights adopted for the different quasar SEDs.

At this point the number of sources under consideration is $3.6\times10^5$. If the colour cut in item iv) is not applied, the number of sources is $8\times10^6$.

These selection criteria are very broad, with the goal of maximising the discovery space of HZQs. The only explicit colour cut is in item 4. It is common to apply a higher S/N threshold than we have in item 1. The very low S/N threshold applied is possible because the BMC method utilises the photometric uncertainties in an optimal way.
One $z>6.4$ quasar in VIKING, listed with $z=6.51$ by \citet{Chen2017},
but revised to $z=6.44$ by \citet{Decarli2018},
lies above, but close to the $Z-Y$ cut. This indicates that the colour cut lies close to $z=6.5$. 
This source is also excluded by the colour cuts applied by \citet{Venemans2013}, described below.
Our cut on $Z-Y$ is slightly redder than that used by \citet{Venemans2013},
which we found helped reduce the number of false positives in our search with BMC.

As detailed below, for sources identified by the BMC method as candidate quasars, we apply an additional cut $\chi^2_\mathrm{min}<11$. The quantity $\chi^2_\mathrm{min}$ is the goodness of fit of the quasar model that provides the best fit to the photometry. 

\subsubsection{Bayesian model comparison method}
\label{sec:bmc}

We use an extension of the BMC technique 
proposed by \citet{Mortlock2012}. 
The extended method is additionally described in detail by \citet{Barnett2019}, so we only recap the main points here.

The method works by determining a `weight' for each population under consideration, 
with a population posterior probability given by the ratio of those weights.
In our search for quasars, denoted $q$,
where we have assumed two contaminating populations, 
MLTs, $s$, and ETGs, $g$,
we define the probability that a source is a quasar, \pq, given photometric data $\bm{d}$ as:
\begin{equation}
\label{eq:pq}
\pq \equiv
p\left(q\given \bm{d}\right) = 
\frac{W_\mathrm{q}\left(\bm{d}\right)}
{W_\mathrm{q}\left(\bm{d}\right)+W_\mathrm{s}\left(\bm{d}\right)+W_\mathrm{g}\left(\bm{d}\right)}.
\end{equation}
The individual weights for a population
are calculated by combining all available photometric data for a source,
with the surface density of the population, which serves as the prior. 
For a given source, 
a particular population weight measures the relative probability 
that the source would have the measured photometry in all bands, 
characterised by the model colours for that population, 
weighted by surface density as a function of apparent magnitude. 
\citet{Mortlock2012} applied the method to the case of two populations: quasars and M stars;
here we have extended the method to three populations. 
Additionally, the MLT population is now divided into a set of sub-populations, 
namely the individual spectral types from M0 to T8.
This approach to the cool dwarf population is similar to that of \citet{Pipien2018},
who developed models for each spectral type \myrange{L0}{T9}
in a search for high-redshift quasars 
in the Canada-France High-$z$ Quasar Survey in the Near-Infrared.

As noted above, for the search we only used one quasar type \---\ the `typical' quasar defined by \citet{Barnett2019}, with standard continuum slope and average line strength. But quasars have a range of spectral properties, so the search will be optimal if this range is represented in the search, i.e., like for the cool stars, we should divide the quasars into sub-populations. The optimal approach would have been to divide the quasars into the nine spectral types, with weights that characterise the relative numbers (see Table 5 in \citealt{Barnett2019}). The use of only one quasar type was an unfortunate consequence of meeting the deadline for applying for 8\,m spectroscopy of candidates, and then the candidate list was locked in. While this means that the search was not quite optimal, the consequences of this are not very significant. The selection function for the search we have undertaken has been correctly computed, i.e., we assume that the actual quasar population has a range of spectral types. The only consequence of the choice made is that the survey goes slightly less deep than the optimal search. Based on calculations presented in \citet{Barnett2019}, we estimate that the optimal search would increase the yield by 20 per cent, i.e., possibly one more quasar would be found.

Explicitly, individual weights for each type, $W_t\left(\bm{d}\right)$, are given by
\begin{equation}
\label{eq:gnrlweight}
W_t\left(\bm{d}\right) =
\int \rho_t\left(\bm{\theta}_t\right)\;
p\left(\bm{d}\given \bm{\theta}_t,t\right)\;
\diff\bm{\theta}_t,
\end{equation}
where $\bm{\theta}_t$ is the set of parameters describing each population. 
The two terms in the integral in Equation~\ref{eq:gnrlweight} are respectively 
the surface density function, 
and a Gaussian likelihood function based on model colours,
which is written in terms of (linear) fluxes (as opposed to magnitudes).

The candidates are ranked on $\pq$, and the candidate list is defined as all objects with $\pq$ larger than some threshold value.
The chosen threshold value of $\pq$ effects a balance between contamination and completeness. 
A value $\pq > 0.1$ worked well for the UKIDSS LAS high-redshift quasar survey \citep{Mortlock2012}.
However, we found the number of candidates rises steeply for VIKING
as the threshold is lowered from $\pq = 0.15$ to $\pq = 0.1$,
with only a small associated change in the simulated selection function, and hence the predicted quasar yield.
The implication is that the lower \pq~threshold
simply allows more contaminants into the sample, to no significant benefit.
Therefore in this work we select candidates with $\pq>0.15$.

In the Appendix, Sect.~\ref{sec:PoS}, we detail a slight modification to the selection procedure
which we applied to sources which have both a primary and secondary entry in VIKING as a result of the VISTA observation strategy
(i.e., the source is duplicated in the catalogue).

\subsection{Candidate selection using other methods}
\label{sec:other}

\subsubsection{SED fitting}
\label{sec:csq}

The second method that we consider is minimum-$\chisq$ SED fitting.
\citet{Reed2017} applied such a method to a combination of DES, VHS and WISE data,
discovering eight bright ($z_{\rm AB} < 21.0$) $z>6$ quasars, 
including one source with $z=6.50$. 
 This method is applied to the same sample that the BMC method is applied to, i.e., after the initial cuts enumerated in Sect.~\ref{sec:bmc}.
For the sake of making an explicit comparison we have attempted to reproduce the methods of \citet{Reed2017} as closely as is possible, given the different datasets. It is likely that some improvements (in terms of contamination and completeness) could be made by making adjustments to the cuts in the parameter space they use. Nevertheless we found that the chosen cuts were sensible when plotting the data in that parameter space. The method was also compared to BMC in \citet{Barnett2019}, and in that study we found that there was little scope for significantly improving the depth achieved by the method. For these reasons, although we have not attempted to optimise the SED fitting method for the VIKING dataset, we feel the comparison as presented provides a useful representation of the relative strengths and weaknesses of the method compared to BMC.

The method works by fitting the full range of contaminant model SEDs, and the nine quasar spectral types, 
to the measured fluxes of a source, 
minimising the reduced $\chisq$ value, $\chisq_{\rm red}$ i.e. the value of $\chisq$ divided by the number of degrees of freedom\footnote{When we use reduced $\chisq$ in this paper there is always an explicit subscript i.e. $\chisq_{\rm red}$}.
We calculate the $\chisq_{\rm red}$ value for a given model SED $m$ as follows:
\begin{equation}
\chisq_{{\rm red,}m} = \frac{1}{N_b - 2} 
\sum_{b}^{N_b}
\left(\frac{\hat{f}_b - s_{\rm best}f_{m,b}}{\hat{\sigma}_b}\right)^2,
\label{eq:chisq}
\end{equation}
where $\hat{f}_b$ and  $\hat{\sigma}_b$ are the measured flux and its uncertainty in band $b$, $f_{m,b}$ is the (unnormalised) model SED flux in band $b$,
and $s_{\rm best}$ is the normalisation that minimises $\chisq$.
We have $N_\mathrm{b} - 2$ degrees of freedom as there are two parameters under consideration: 
the normalisation of a single model, and the particular model being fitted, selected from a range of models \citep[e.g.,][]{Skrzypek2015}.
That is to say, for the quasars and early-type galaxies the second parameter is redshift, $\Delta z = 0.05$, while for the MLT dwarfs, which form a continuous sequence, the second parameter is spectral type.

We use the model colours detailed in the Appendix, Sect.~\ref{sec:pops}, to produce quasar and contaminant SEDs,
and fit them to the fluxes of each source, following Eq.\,(\ref{eq:chisq}).
We keep the single best fitting quasar $(q)$ model and contaminant $(c)$ model, 
with respective $\chisq_{\rm red}$ values $\chisq_{{\rm red,q(best)}}$
and $\chisq_{{\rm red,c(best)}}$. 
Following \citet{Reed2017}, 
we apply two cuts to the $\chisq_{\rm red}$ values to retain a source (see Figure 15 of that work).
We firstly require $\chisq_{{\rm red,c(best)}} > 10$, 
i.e., the data are a bad fit to all contaminant models. 
We additionally require the ratio $\chisq_{{\rm red,c(best)}}/\chisq_{{\rm red,q(best)}} > 3$,
i.e., the data are fit substantially better by a quasar SED than any contaminant model.

\subsubsection{Colour cuts}
\label{sec:cc}

The final method we use in this paper is the `conservative' set of colour/magnitude cuts
used to select quasars from 332\,deg$^2$ of VIKING by \citet{Venemans2013}. The selection criteria, transformed from the AB photometric system used by them to the Vega system used here, are the following: 

\begin{enumerate}

  \item $\zy > 1.35$, or undetected in $Z$ (VIKING).
  
  \item $\sn_Z < 3$ or $ \zy > 1.3 + 0.75\,(\yj)$.
  
  \item $-0.2 < \yj < 0.8$.
  
  \item $0.7 < \yk <  2.5$, or unmeasured in $\ks$.
  
  \item $\sigma_Y(\textrm{VIKING}) < 0.15$. This requirement limits the search to redshifts $z<7.5$, in contrast to the other two search methods used here.
  
  \item Not on detector 16. 

\end{enumerate}

A source must satisfy all criteria to be accepted. 
In their search, \citet{Venemans2013} also undertook their own aperture photometry on the images, and additionally applied the above cuts to the repeat photometry. Since we do not have access to their software,  we use the list-driven photometry to emulate this process, i.e., we apply the cuts first to the standard VIKING photometry, then additionally to the list-driven photometry. In the first stage we also apply the morphological cut $\mathrm{pGalaxy}<0.95$ used by  \citet{Venemans2013}.

\section{Results: $\lowercase{z}>6.5$ quasar candidates}
\label{sec:search}

We present the BMC candidate list, and the results of follow-up observations, in Sect.~\ref{sec:bmcsearch}.
We did not discover any new quasars; however, the four known quasars in the VIKING area are easily recovered.
In Sect.~\ref{sec:otherlists} we outline the candidate lists produced using the other two methods.
As noted above these are only for the purposes of comparing the efficiency and depth of the three different selection methods.

\subsection{Results from Bayesian model comparison}
\label{sec:bmcsearch}

\begin{table*} 
	\centering
	\advance\leftskip-3cm
	\advance\rightskip-3cm
	\caption{VIKING HZQ candidates from the BMC method. Follow-up imaging ($g$ or $r$) and spectroscopic (FORS2/FIRE) observations are listed in the final column. Here `Legacy' means the DESI Legacy Survey.
		The asterisked probabilities were calculated on the basis of combining 
		primary and secondary VIKING photometry for those sources (see Appendix Sect.~\ref{sec:PoS}). The quantity $\chi^2_\mathrm{min}$ has three degrees of freedom.}
	\label{tab:bmccands}
	\begin{tabular}{lcclrll}
		\hline \hline \\[-2ex]
		& $\alpha$ & $\delta$ & \pq & $\chi^2_\mathrm{min}$ & $z_\textrm{best}$ & notes \\[0.5ex]
		\hline \\[-2ex] 
	Known HZQs & 01:09:53.12 & --30:47:26.3 & 0.97 & 0.6 & 7.00  & $z = 6.8$, \citet{Venemans2013}\\
		& 03:05:16.91 & --31:50:55.9 & $1.00^*$ & 2.1 & 6.55  & $z = 6.6$, \citet{Venemans2013} \\
		& 10:48:19.08	& --01:09:40.2 & 1.00 & 4.5 & 6.60  & $z = 6.6$, \citet{Wang2017}\\
		& 23:48:33.33	& --30:54:10.2 & $1.00^*$ & 2.3 & 7.00  & $z = 6.9$, \citet{Venemans2013}\\
		\hline \\[-2ex] 
		Candidates & 00:03:51.28	&--31:24:00.2	& 0.21	&1.9&	6.65	&	FORS2 \\ 
		&00:27:57.62	&--30:02:19.7	&0.20	&1.9&	6.45&	Legacy $r$	 \\
		&00:45:39.65	&--34:28:02.3	&0.17	&0.1&	7.45&	FORS2 \\ 
		&00:58:42.88	&--28:52:06.1	&0.42	&3.3&	6.50&	Legacy $r$	 \\
		&01:00:23.22	&--28:55:36.5	&0.19	&1.5&	6.55&	Legacy $r$	 \\
		&01:21:51.96	&--28:00:18.4	&0.99	&5.8&	6.40&	Legacy $r$	 \\
		&01:40:12.32	&--27:54:04.7	&0.17	&1.3&	6.50&	Legacy $r$ \\ 
		&01:46:30.52	&--30:11:51.0	&0.63	&0.4&	7.40&	FORS2 \\ 
		&02:19:41.29	&--27:25:33.8	&0.38	&5.9&	7.00&    Legacy $r$, FORS2\\ 
		&03:13:16.83 &--30:59:20.9	&$0.77^*$	&6.9&	6.45&	Legacy $r$ \\ 
		&08:59:02.86	&--01:36:02.5	&0.25	&2.4&	6.45&	Legacy $r$ \\ 
		&12:05:47.13	&+01:52:54.3	&$0.37^*$	&2.1&	6.50	&   Legacy $r$ \\ 
		&14:08:11.22	&--02:44:39.3	&0.29	&1.6&	6.60	&	FIRE \\
		&22:02:09.61	&--28:19:51.0	&0.27	&1.1&	7.50&	FORS2 \\ 
		&22:25:11.17	&--27:23:29.6	&0.16	&3.6&	6.45&	Pan-STARRS $r$	 \\
		&22:39:54.55 &--27:12:18.1	&0.64	&1.6&	6.45&	ATLAS $r$	 \\
		&23:24:17.94	&--30:12:12.3	&0.52	&5.4&	6.45&	KiDS $r$	 \\
		\hline \\[-2ex] 
		Rejects & 01:13:32.06	&--30:08:45.6	&0.91&	18.4&	6.50	& Legacy $r$ \\
		$11<\chi^2_\mathrm{min}<40$&01:27:59.05	&--33:02:57.1	&0.31&	25.1&	6.45	& Legacy $r$	 \\
		&02:16:04.31	&--32:58:58.7	&0.39&	31.4 &	7.00&	FORS2; Fig.~\ref{fig:spectrum_red} \\ 
		&02:16:23.37 &--32:07:40.4	&0.99&	32.3&	6.55&	FORS2 \\ 
		&11:48:27.75	&+02:53:51.8	&0.53&	19.8&	6.45&	Legacy $r$	 \\
		&12:25:39.61	&+02:31:27.8	&0.47&	14.4&	6.45&    Legacy $r$	 \\
		&12:45:09.26 &--01:40:23.3   &0.62&	27.4&	6.50&	ATLAS $g$	 \\
		&12:51:50.09 &+02:50:16.2	&0.98&	35.0&	6.40&	Legacy $r$	 \\
		&14:45:21.04 &+02:00:58.4	&0.99&	36.3&	7.00&	Legacy $r$	 \\
		&14:59:14.63	&--03:21:29.1	&0.20&	22.0&	6.50&	Legacy $r$, FORS2; ELG, Fig.~\ref{fig:spectrum}\\ 
		&23:04:16.06	&--34:52:30.8	&0.20&	24.2&	6.45&	ATLAS $g$ \\
		\hline
	\end{tabular}
\end{table*}

Applying our first cuts to the full VIKING survey (Sect.~\ref{sec:methods}),
we have an initial sample of $3.6\times10^5$ sources, to which we apply the BMC method. This produced an initial list of 349 candidates satisfying $\pq>0.15$. All candidates were then checked on the VIKING images. The majority were eliminated as obvious spurious images arising from, e.g., detector flaws, diffraction spikes, and satellite trails, and only 42 real sources remained. At this stage we were left with a number of candidates that have SEDs very different to the SEDs of quasars.
For each candidate we determined the best-fit quasar SED, by min-$\chi^2$, using all nine quasar types, with redshifts over the range $6<z<10$. For sources with measurements in all five bands $ZYJHK_s$ 
this fit has three degrees of freedom, so we rejected all candidates with $\chi^2_\mathrm{min}>11$. 
For three degrees of freedom, this cut corresponds to a 1 per cent probability,
i.e., in principle we include 99 per cent of quasars with this cut.
We removed 21 sources in this way, leaving 21 candidates.
The nature of the sources eliminated by this cut is discussed later in this section\footnote{ The use of $\chi^2$ at this stage is to measure the goodness of fit. This is distinct from the use of $\chi^2$ in the SED-fitting method, where it is employed for model comparison.}.

At this stage we checked the photometry of the remaining sources in the ALLWISE database (this combines WISE \citep{Wright2008} with NEOWISE \citep{Mainzer2011}), since late T dwarfs would appear bright in the W2 band. Many of the candidates are undetected in ALLWISE, and no sources were eliminated in this way. ALLWISE is of limited utility for searches for HZQs at the depth of VIKING. The final list of 21 candidates is provided in Table~\ref{tab:bmccands}. We also list the 11 rejected sources with $11<\chi^2_\mathrm{min}<40$ (10 rejected sources have $\chi^2_\mathrm{min}>40$ and are not considered further). The Table lists the coordinates (IRCS, truncated, according to IAU convention), the value of $\pq$, the value of $\chi^2_\mathrm{min}$, and $z_\mathrm{best}$, the redshift of the best-fit quasar SED. Ranking the candidates by $\pq$, the four known HZQs are ranked numbers 1, 2, 3, 5. 

\begin{figure}
	\centering
	\includegraphics[width=9cm]{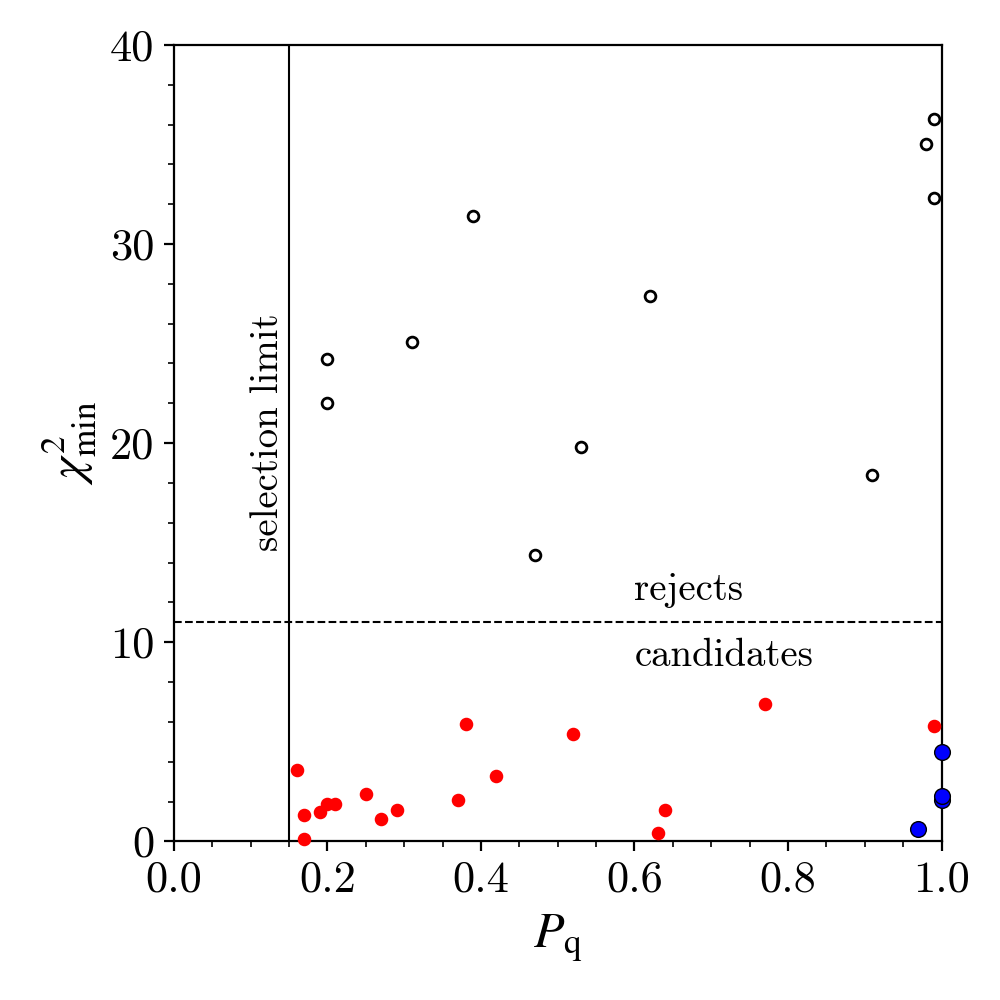}
	\caption{Plot of $\chi^2_\mathrm{min}$ against \pq~for the 21 candidates, and the 11 rejected sources with $11<\chi^2_\mathrm{min}<40$. The four larger blue circles (two overlap) are the four quasars and the red circles are the 17 candidates subsequently shown to not be quasars.
	}
	\label{fig:candidates}
\end{figure}

The 21 candidates as well as the 11 rejected sources with $11<\chi^2_\mathrm{min}<40$ are plotted in Fig. \ref{fig:candidates}. The number of candidates increases extremely rapidly below the selection cut $\pq=0.15$ (see \citet{Mortlock2012}, Fig. 11, for an illuminating plot on this point). It is noticeable that the distribution of $\chi^2_\mathrm{min}$ for the blue and red points is sensible for three degrees of freedom\footnote{ For the 21 blue and red points $\chi^2_\mathrm{min}$ has mean 2.7 and standard deviation 1.9. The expected values, for a larger sample, are 3 and 2.4 respectively.}, but that there is an excess of sources with bad fits, with $\chi^2_\mathrm{min}>11$. 

Since we want a complete sample we must confirm whether or not any of the remaining 17 candidates is a quasar. At redshifts $z>6.5$, flux shortward of the quasar Ly$\alpha$ emission line is almost completely absorbed, with at most some residual transmission redward of Ly$\beta$ which lies at $\lambda>770$\,nm \citep[see, e.g.,][for recent measurements]{Barnett2017}. Therefore if a candidate is a HZQ, there will be negligible flux in the $g$ and $r$ photometric bands, which lie blueward of $\lambda700$\,nm.
(We therefore ignore the possibility of gravitational lensing by an intervening galaxy that
magnifies the quasar image(s) and directly contributes optical flux, see \citealt{Fan2019}.)
We matched our candidates to imaging data from the DESI Legacy Survey,
the VST ATLAS survey \citep{Shanks2015}, the Kilo-Degree Survey \citep[KiDS,][]{deJong2017}, and Pan-STARRS, and eliminated any candidates clearly detected in either of these bands\footnote{In
	principle this additional optical photometry could have been incorporated into the BMC calculation.
	However, for our search insufficient optical data were available at the time the candidate list was finalised. In any case, matching once we have a shortlist is considerably less work than matching complete catalogues before starting the search.}.
The images (survey and photometric band) used to eliminate any candidate are listed in the final column of Table~\ref{tab:bmccands}.

\begin{figure}
	\centering
	\includegraphics[width=9cm]{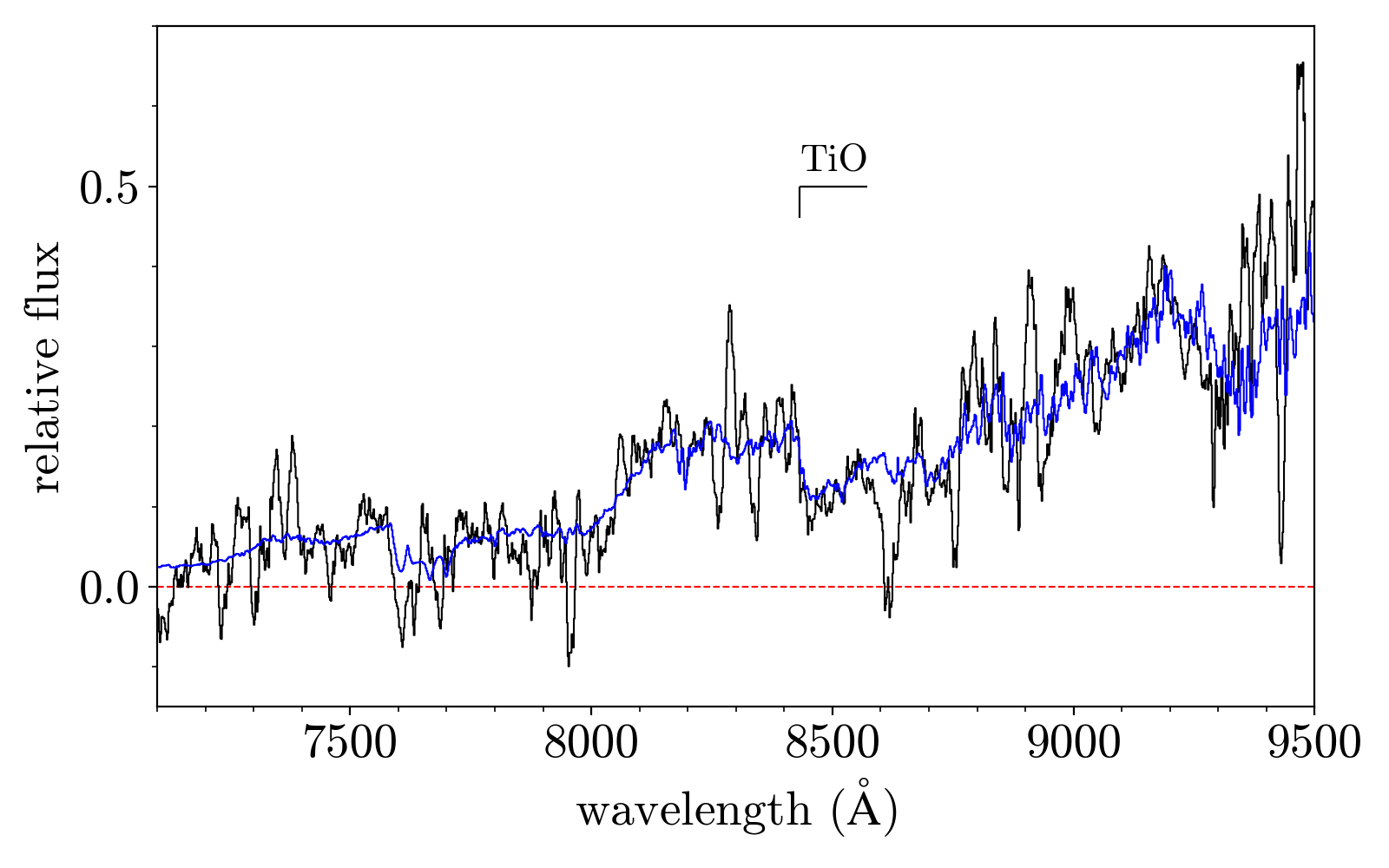}
	\caption{FORS2 spectrum of VIK J0216--3258, which we identify as a late M or early L dwarf. The spectrum has been flux calibrated relatively, and smoothed. Overplotted in blue is the scaled spectrum of the L0 standard 2MASS J0345+2540 from \citet{Kirkpatrick1999}.}
	\label{fig:spectrum_red}
\end{figure}

We obtained spectra to confirm the nature of the remaining five objects.
One of these, VIK J140811.2--024439.4, was observed using Magellan FIRE in June 2017.
The other four candidates were observed with the ESO FORS2 instrument between \nth{11} November 2018 and \nth{18} March 2019, as well as one additional source VIK J021941.3--272533.8 that was only later detected in a $r$ image. None of the six spectra display the continuum break or Ly$\alpha$ emission line characteristic of HZQs. Instead all display continua strongly rising towards longer wavelengths, characteristic of L and T dwarfs. They are similar to the spectrum of the source VIK J021604.3--325858.7, plotted in Fig. \ref{fig:spectrum_red}, that is discussed in the next section. Accordingly these sources were also eliminated. In this way all 17 candidates were eliminated, meaning no new quasars were discovered. By this means, the four previously known quasars now form a complete sample.

\begin{figure}
	\centering
	\includegraphics[width=9cm]{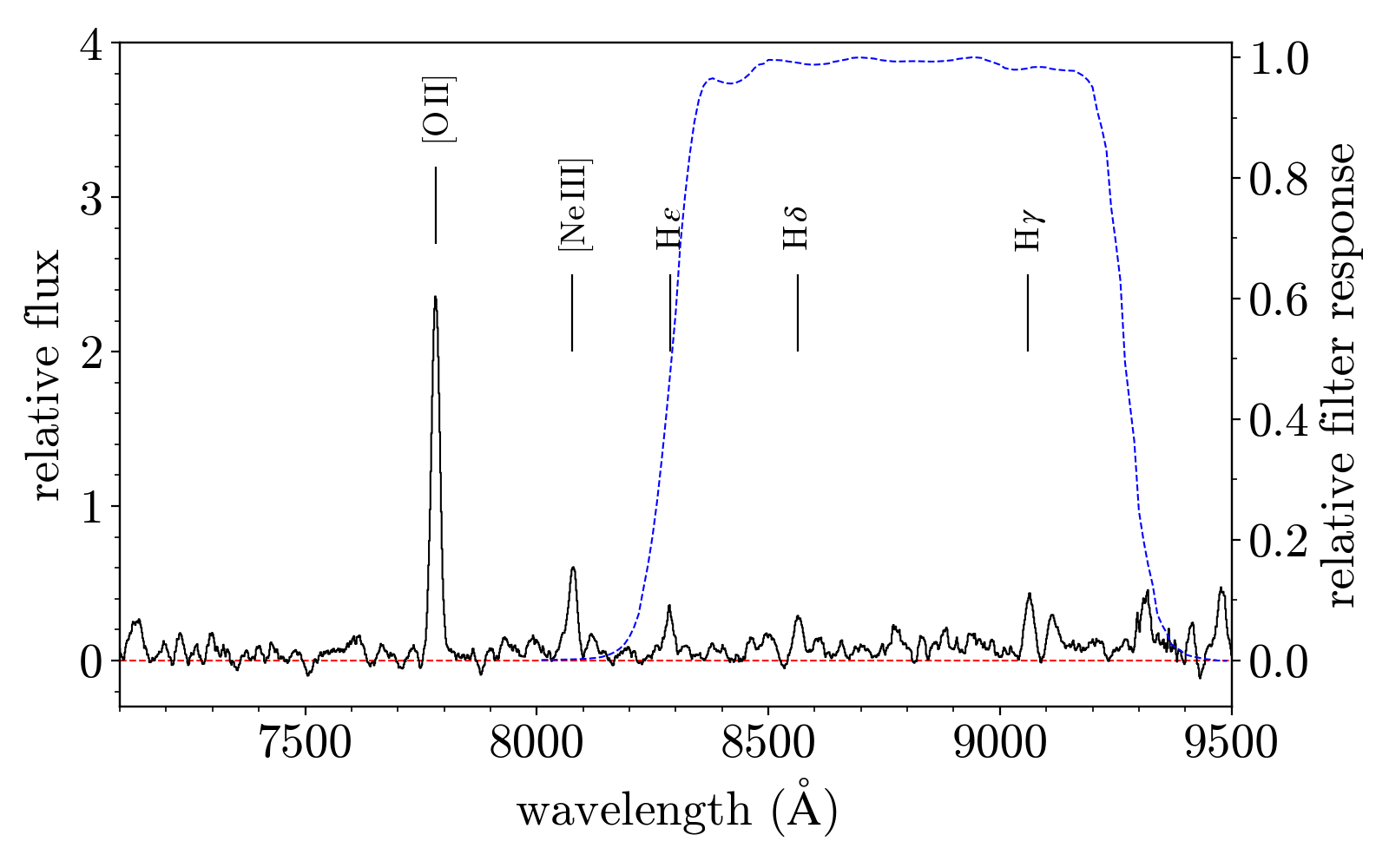}
	\caption{FORS2 spectrum of VIK J1459--0321, a $z=1.087$ emission line galaxy. The spectrum has been flux calibrated relatively, and smoothed. The (relative) transmission profile of the $Z$ filter is plotted as the blue dashed line, showing that the strong [OII] line lies outside the filter.}
	\label{fig:spectrum}
\end{figure}

\subsubsection{The nature of $\chi^2_\mathrm{min}>$ 11 sources}
A number of candidates have acceptable values of $\pq>0.15$ but the best fit quasar model is a very bad fit, with $\chi^2_\mathrm{min}>11$. This is not a shortcoming of the BMC method {\em per se}. From the range of models (i.e, quasars and contaminants), the BMC method optimally selects the model that best explains the data. A bad fit of the best model suggests
that the object is not a member of any of the three populations considered, and may indicate the presence of another contaminating population. The BMC method itself does not measure goodness of fit, which is why we assess the goodness of fit using $\chi^2$. 

In Table~\ref{tab:bmccands} we list the objects with $11<\chi^2_\mathrm{min}<40$. To investigate the nature of this population we again matched the sources to the deep optical surveys previously cited. Most are detected in the $r$ band, and these are considered further below. Two sources were not detected in $g$ or $r$ and we obtained spectra of these. Their spectra are similar to the spectra of the red objects discussed in the preceding section. The spectrum of the source J021604.3--325858.7, $J=20.1$, is plotted in Fig. \ref{fig:spectrum_red}. The wavelength of the sharp dip near the centre of the spectrum matches to the 8432\AA\, bandhead of TiO. The overall shape of the spectrum, the TiO absorption, and the dip near 9300\,\AA\,identified as H$_2$O, are characteristic of a source near the M/L boundary, and as shown in the figure the spectrum is satisfactorily fit by the L0 standard. The reason for the large value of $\chi^2_\mathrm{min}$ is not clear, but conceivably this source and the other one undetected in optical imaging are subdwarfs, but this cannot be confirmed without much longer integrations. 

We also targeted a source detected in $r$ with one of the smallest values of $\chi^2_\mathrm{min}$ in this group. The spectrum of this source, VIK J145914.6--032129.11 (VIK J1459--0321), is plotted in Fig.~\ref{fig:spectrum}. It is an emission line galaxy (ELG) with $z=1.087$. 
The colours of this source are plotted in Fig.~\ref{fig:colourcolour}, where it can be seen that the object is very blue in $Y-J$. Given the redshift of the source, the emission lines H$\beta$ and [OIII]$\lambda\lambda$4959,5007 lie in the $Y$ band, but off the red end of the spectrum plotted in Fig.~\ref{fig:spectrum}. From the spectrum it is clear that the Balmer lines are weak, so the very blue $Y-J$ colour must be caused by very strong [OIII]$\lambda\lambda$4959,5007 emission. The strength of the high-ionisation [NeIII] line compared to the Balmer lines suggests this object is likely to be an AGN. Also plotted in Fig.~\ref{fig:colourcolour} are the average colours, and scatter, of the nine sources in the high$-\chi^2_\mathrm{min}$ reject category that are visible in $g$ or $r$. These are consistent with the colours of VIK J1459--0321, suggesting that they may all be ELGs at similar redshifts. While these objects are easily distinguished from quasars, by their high $\chi^2_\mathrm{min}$ values, similar sources with somewhat weaker emission lines might have less blue $Y-J$ colours, and so smaller values of $\chi^2_\mathrm{min}$. It is possible that some of the actual candidates, i.e., objects with $\chi^2_\mathrm{min}<11$, detected on $r$, are also ELGs. 

This population of ELGs, with very strong [OIII]$\lambda\lambda$4959,5007 emission, is discussed in the literature \citep[e.g.,][]{Atek2011,Hayashi2018}, and has been noted as a contaminant in the very deep Subaru surveys for HZQs by \citet{Matsuoka2019}. It appears we are seeing the bright tip of this population in VIKING. As described in the {\em Euclid} paper we made an explicit search of the deep COSMOS data \citep{Laigle2016} for an additional contaminating population, failing to find any such sources over the 1.5\,deg$^2$ of the COSMOS field, but the area covered was evidently too small to capture this population. The {\em Euclid} wide survey will cover an area 10\,000 times larger. Therefore further analysis and better characterisation of this ELG population is needed, to feed into HZQ selection with \textit{Euclid}. It may be that the much better imaging quality of \textit{Euclid} compared to VIKING
will allow the ELG population to be eliminated from quasar searches as extended sources, or that the complementary ground-based optical imaging data will be deep enough to detect these sources, in $r$ or $i$, and so eliminate them in that way.

A further known contaminant in $z>6.5$ quasar surveys are FeLoBALs \citep{Hall2002},
although none of the objects for which we obtained a spectrum in this work is an example.
FeLoBALs are best eliminated by deep photometry in the optical bands, 
and it is conceivable that some of our objects detected in $r$ fall into this category.

\subsection{Candidate lists from other methods}
\label{sec:otherlists}

We applied the SED fitting criteria to the same initial sample of $3.6\times10^5$ sources as in Sect.~\ref{sec:bmcsearch}.
After checking the VIKING images of all candidates we were left with only two candidates, 
of which one is the $z=6.9$ quasar listed in Table~\ref{tab:bmccands}, and the other is one of the BMC candidates, which we excluded on the basis of a DESI Legacy Survey detection.
The other three known quasars are excluded as the photometry is well-enough fit by a contaminant model. This implies that the SED fitting method is very efficient, but only picks out the most obvious sources. 
In our analysis for {\em Euclid} \citep{Barnett2019}, we investigated whether tuning the selection parameters for SED fitting could improve the completeness.
However, in that work we found relaxing the cuts slightly resulted in a threefold increase in contamination,
for only a 10 per cent increase in the predicted number of quasars found.
We concluded that it was difficult to improve the depth significantly without the number of candidates rapidly rising. In their more recent search \citet{Reed2019} used different SED-fitting criteria compared to \citet{Reed2017}, extending the region of parameter space searched. However, because they have not obtained spectra of all candidates a quantitative comparison of the two selection methods is not possible. In the newer search they found three new HZQs. Two of these satisfy the selection criteria of \citet{Reed2017}, while the third lies outside but very close to the cut $\chisq_{{\rm red,c(best)}}/\chisq_{{\rm red,q(best)}} > 3$.

Using the colour cuts, and after checking all candidates in the VIKING images, we are left with a total of 199 good HZQ candidates. We recover the four known quasars as well as five BMC candidates listed in Table~\ref{tab:bmccands}
(all of which have detections in the $r$ band). 
We find a further 124 candidates which are not selected by the BMC algorithm because $\pq<0.15$.
Finally, there are 66 additional candidates which were not checked using BMC, 
as they lie bluer than the $Z-Y$ cut applied to produce the initial sample used in the BMC method. 
The surface density of the 199 candidates, from 977\,\dsq, is 0.20\,deg$^{-2}$. 
\citet{Venemans2013}, selected 43 candidates from 332\,\dsq, or 0.13\,deg$^{-2}$, so our methods are reproducing theirs reasonably closely.

Considering the colour range in common, the BMC method produces 17 false positives. Colour cuts produce 129 candidates in addition to the known quasars. Considering the high probabilities of the confirmed quasars it can be assumed that nearly all the 129 candidates are false positives, so on this basis we find colour cuts produce eight times as many false positives as the BMC method.

\section{Selection functions}
\label{sec:simulations}

\begin{figure*}
	\centering
	\subfloat[BMC]{%
		\includegraphics[trim=10 0 100 0,clip,totalheight=0.21\textheight]{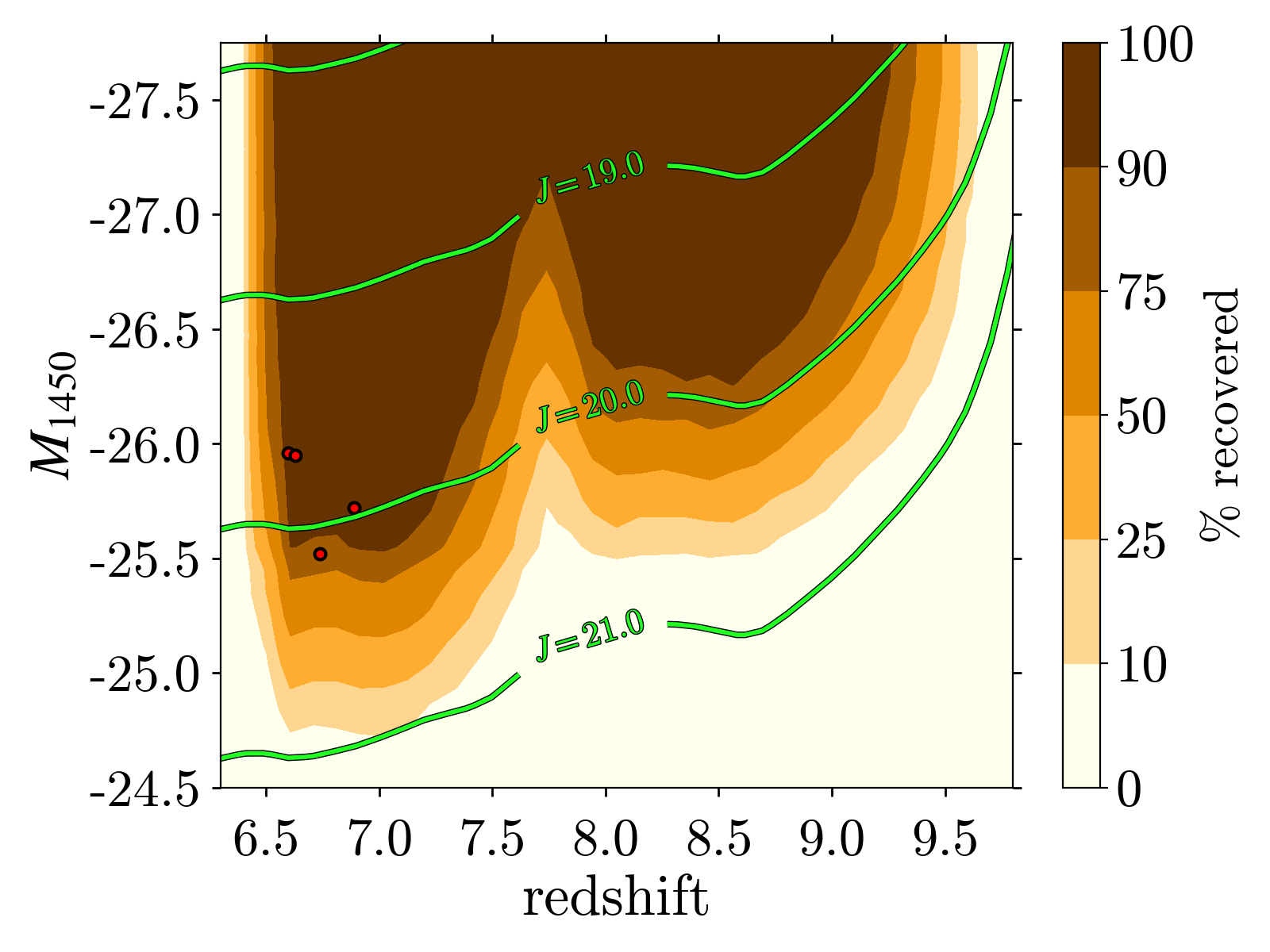}%
		\label{fig:ctrBMC}%
	}\,
	\subfloat[$\chisq$ fitting]{%
		\includegraphics[trim=10 0 100 0,clip,totalheight=0.21\textheight]{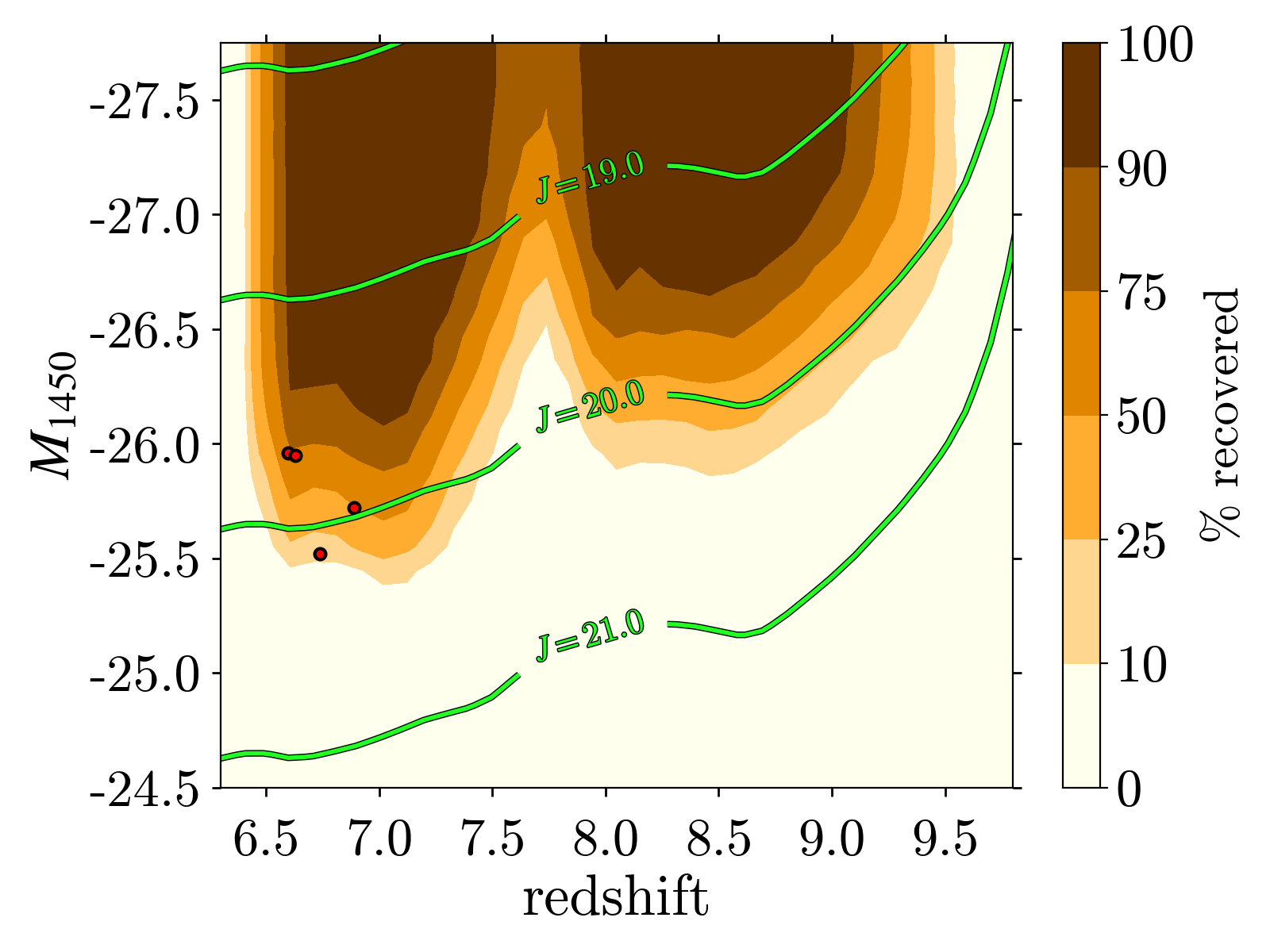}%
		\label{fig:ctrCSQ}%
	}\,
	\subfloat[Colour cuts]{%
		\includegraphics[trim=10 0 0 0,clip,totalheight=0.21\textheight]{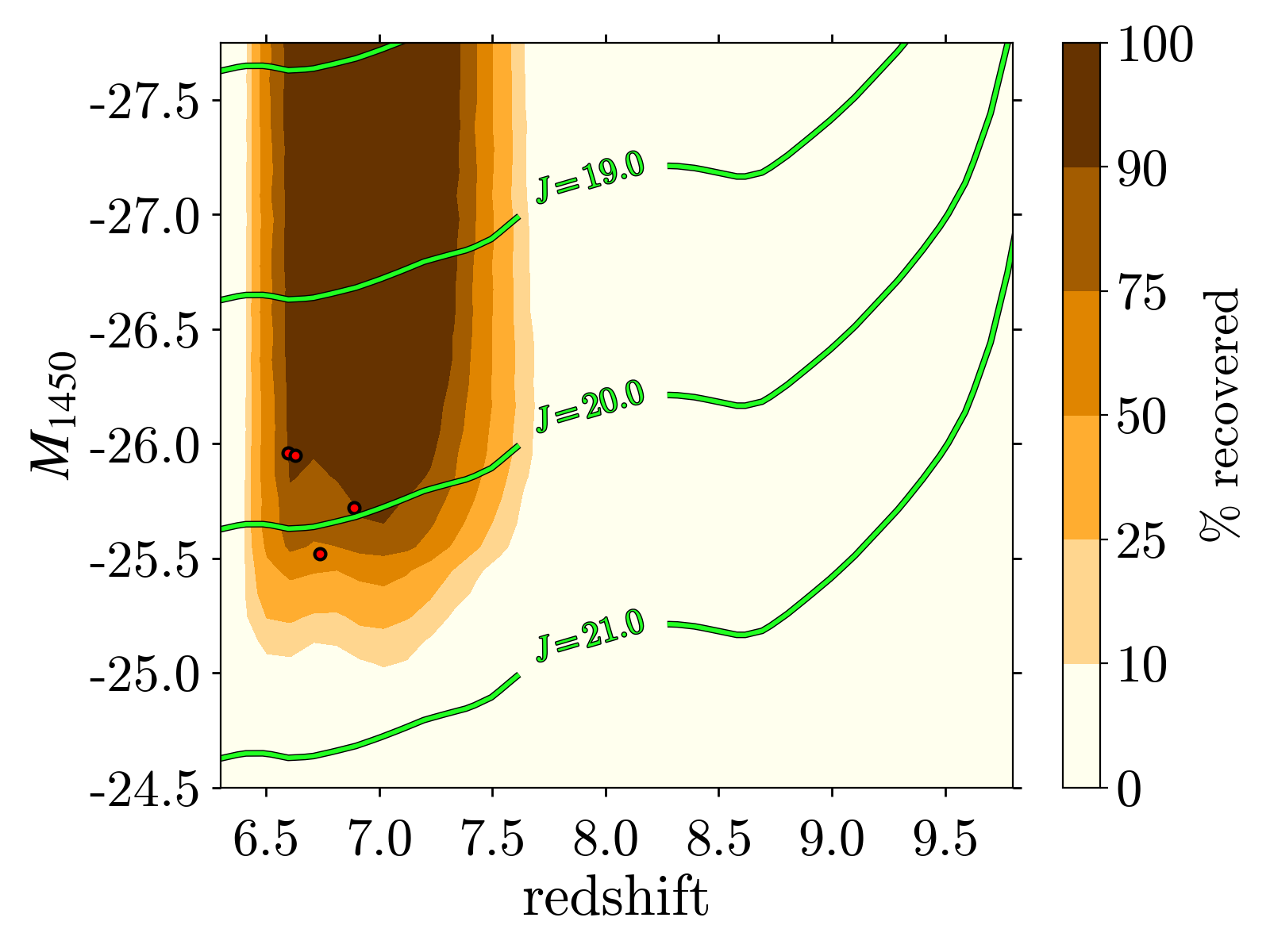}%
		\label{fig:ctrCCS}%
	}
	
	\caption{Selection functions for the three different selection methods.
		Circles indicate the four published VIKING HZQs \citep{Venemans2013,Wang2017}.
		Contours of constant apparent magnitude, computed using  $k$-corrections determined for an average quasar SED, are indicated in green.}
	\label{fig:contour}
\end{figure*}

\begin{figure}
	\centering
	\includegraphics[width=9cm]{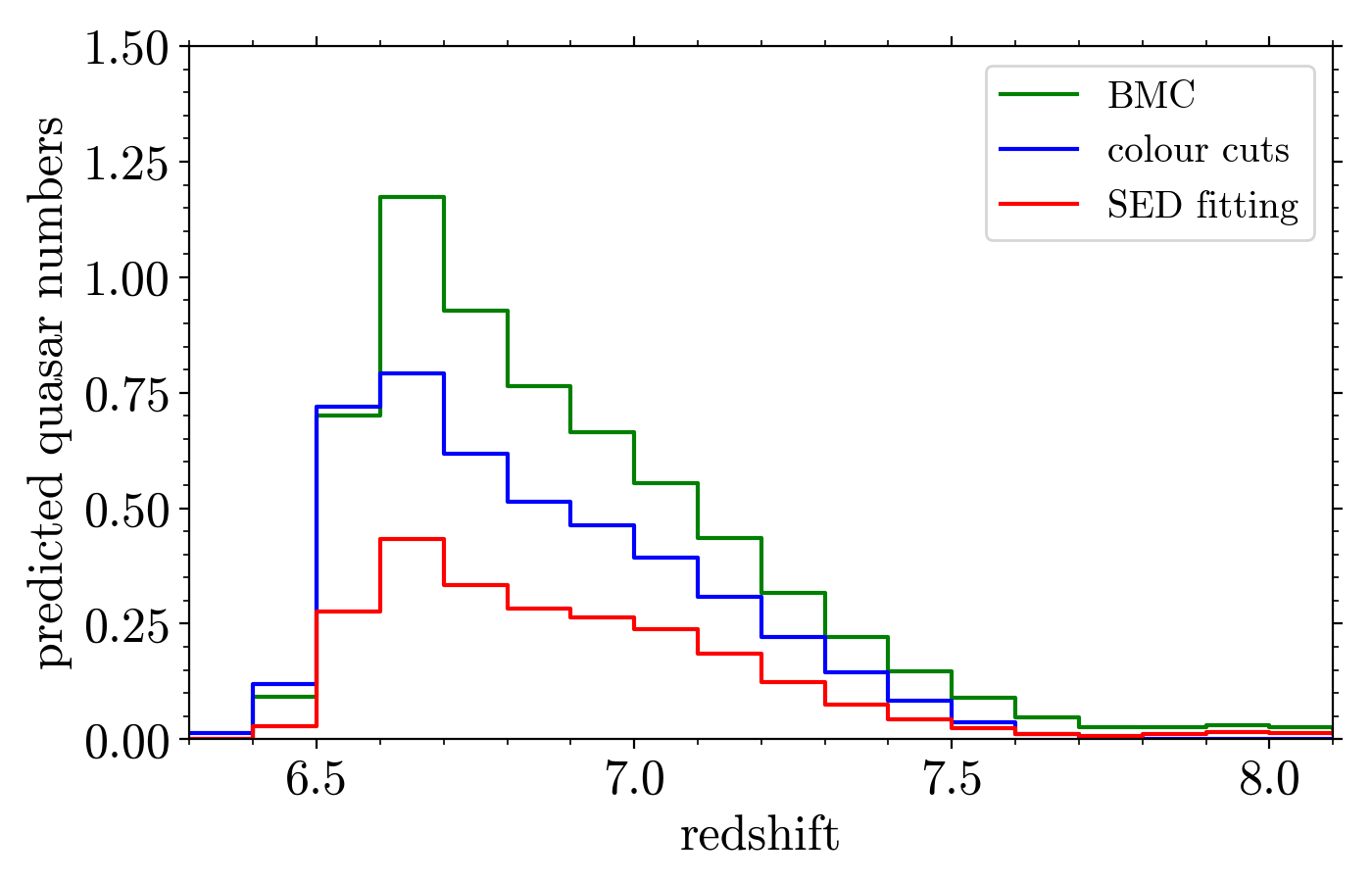}
	\caption{Predicted HZQ yield as a function of redshift for each selection method, 
		produced by multiplying the selection functions in Fig.~\ref{fig:contour} by the \citet{Jiang2016} QLF,
		evolved beyond $z=6$ using $k=-0.78$.}
	\label{fig:counts}
\end{figure}

In this section we compute the quasar selection functions for the three selection methods, i.e., we compute the probability that a quasar of given absolute magnitude and redshift and spectral type, would be selected by each method.
We start with a grid in luminosity/redshift space,
over which we simulate large numbers of quasars. Although the selection algorithm employs only a single quasar spectral type, one with typical properties, quasars in reality have a range of emission-line strengths, and continuum slopes. We employ nine different spectral types, corresponding to all combinations of three different line strengths and three continuum slopes. Further details of the models are provided in \citet{Barnett2019}. We measure the detection probability over the full range of spectral types. The final selection function weights the detection probabilities by the assumed fractions of the different spectral types (Table 5 in \citet{Barnett2019}). 

We produce realistic list-driven fluxes for these sources
using our model colours (Appendix, Sect.~\ref{sec:mq}),
and add Gaussian noise based on the distribution of frameset depths in each band (Fig.~\ref{fig:depths}).
We additionally simulate the  process of detection in the $J$ band,
which determines whether a real source will appear in the list-driven catalogue.
The way the detection process is modelled is described in the Appendix, Sect.~\ref{sec:casu}. This is relevant for the BMC method, which reaches very deep.

We determine the final quasar selection functions by recording the fraction of simulated HZQs 
that pass the selection criteria for each method, considering every step in the selection process, i.e., following the procedures detailed in Sect.~\ref{sec:selection}.
The three selection functions are plotted in Fig.~\ref{fig:contour}.

There is a sharp cut-off in sensitivity to quasars below $z=6.5$ for all three methods.
Colour cuts are insensitive to quasars at $z\gtrsim7.6$, 
beyond which the \lya break is sufficiently redshifted that HZQs become too faint in the $Y$ band to be selected.
By contrast there is no $Y$-band \sn requirement for the other two methods
(a source with a $4\,\sigma$ flux measurement in any of $YH\ks$ will be accepted),
and quasars could in principle be recovered up to $z\sim9.3$.
For the BMC and SED fitting methods there is a noticeable decline in selection efficiency 
over the redshift interval $7.5<z<8.0$.
Over this redshift range, HZQs fainter than $J\sim19$ begin to be misclassified as MLTs,
as a result of the very similar colours of the two populations.

Over the redshift range in common to the three methods, the BMC method reaches deepest, followed by colour cuts, and then SED fitting.
To quantify the relative performance of the three HZQ selection methods,
we integrate the QLF over the quasar selection functions in Fig.~\ref{fig:contour},
to determine a `predicted yield' of HZQs using each method.
We use the \citet{Jiang2016} QLF measured at $z=6$,
and evolve it in redshift using the value $k=-0.78$ measured by \citet{Wang2019}, assuming this value applies
at all redshifts $z>6$. We plot the predicted numbers in redshift bins in Fig.~\ref{fig:counts}. The total counts are 6.4, 2.4, 4.4 for the BMC, SED fitting, and colour cuts methods. Although the BMC and SED-fitting methods have sensitivity to $z>8$ quasars, in practise this is not of great interest as the predicted numbers are negligible for the assumed luminosity function evolution.
The tally of four detected quasars using BMC is effectively a $-1\,\sigma$ deviation from the predicted counts\footnote{To further check the consistency of the BMC selection function and VIKING HZQ sample we draw $1\times10^5$ fake quasar samples from the product of the selection function and the QLF. We calculate the log likelihood of each \citep{Marshall1983}. The log likelihood of the VIKING sample is consistent with the resulting distribution.}.
The implication of the results is that the rate of decline of the luminosity function with redshift is slightly steeper than found by \citet{Wang2019}. In a future paper we will combine our dataset with that of \citet{Wang2019}, to carry out a full analysis of the redshift evolution at $z>7$.

To quantify the relative depths of the three selection methods we consider the redshift range $6.6<z<7.0$. We sum the predicted counts over this redshift range for each method. We then similarly integrate the evolving luminosity function, assuming 100 per cent completeness, to find the depth that matches the counts for each method. 
The effective depths of the BMC, SED fitting, and colour cuts methods are respectively
$M_{1450} = -25.0,-25.7,-25.3$. This implies that the BMC method reaches 0.3\,mag. deeper than colour cuts,   
and 0.7\,mag. deeper than SED fitting.


\section{Contaminant simulations}
\label{sec:mltell}

As a test of the modelling process, and therefore of the reliability of the selection functions, we created a synthetic survey using the population modelling and selection apparatus already described. We ran our selection methods on the synthetic survey and compared the numbers of synthetic candidates to the numbers of actual candidates.
Our models for the contaminants specify the surface density as a function of apparent magnitude and colour. The synthetic survey accounts for 
the varying depths across the VIKING survey
by simulating sources for every frameset.
We simulate sources down to one magnitude fainter than the $5\,\sigma$ limit of each frameset in order to correctly model the possibility that faint sources are scattered by flux errors to brighter than the detection limit.
For both contaminating populations we use the model colours to 
determine true fluxes in each band, and then add Gaussian noise based on the frameset depth.
We discard sources that are not detected in the $J$ band,
and in the case of ETGs we reject sources on the basis of the morphology cut (Appendix Sect.~\ref{sec:mg}).
The resulting synthetic survey contains $\sim 1.5\times10^6$ MLTs,
and $\sim 4\times10^5$ ETGs (the large majority having been removed by the morphology cut).
We gauge the predicted contamination by applying our selection methods to this sample.

For the BMC method we predict contamination by 29 sources, as compared to the 17 contaminants found (effectively a $-2\,\sigma$ fluctuation), and for SED fitting we predict one contaminant which matches the one found. These numbers are in very close agreement, providing confidence in the models used and therefore in the accuracy of the computed selection functions.
However, we note that our idealised simulations 
do not incorporate the possible ELG population discussed in Sect.~\ref{sec:bmcsearch}, and that at present it is unclear what fraction of the contaminants are ELGs.

We cannot fully simulate the colour cuts method using simulations. Recall (Sect.~\ref{sec:cc}) that \citet{Venemans2013} required an object to meet their selection criteria both in the catalogue data and in their own repeat photometry. While we could emulate this in the candidate selection, we cannot do this in the simulation. Because of this, the simulation is expected to overestimate the number of candidates. In the actual data we selected 199 candidates using colour cuts. Applying the colour-cut criteria to the simulations we select 472 HZQ candidates. There is a discrepancy of a factor 2.4. Although we predicted an excess, this difference is substantial. Nevertheless, the context of this estimate is that over the parameter space searched, ETGs outnumber HZQs by a factor larger than $10^4$. So here we are attempting to model only the extreme tail of the ETG distribution. Coupled with the complication of accurately modelling the MCS cut we consider this a satisfactory prediction of the number of contaminants. We conclude that these results again indicate the models are providing a good representation of the contaminant populations.

We have shown in this section that we have a good understand of the dominant contaminating populations. This implies that the selection algorithm is nearly optimal.  

\section{Conclusions}
\label{sec:end}
In this paper we have presented a search for redshift $6.5<z<9.3$ quasars over 977\,deg$^2$ of the VIKING survey.
We have exploited a new list-driven dataset,
which provides fluxes and uncertainties for all $J$-detected VIKING sources,
in all available photometric bands $ZYJH\ks$. We searched the database using a modification of the BMC method of \citet{Mortlock2012}, and produced a sample of 21 candidate quasars $z>6.5$. This candidate list includes the four previously discovered $z>6.5$ quasars in this field. 

We have followed up the additional 17 candidates and confirmed that none are quasars. The sample and the selection criteria define a complete sample and allow us to compute the survey selection function for VIKING for the first time. The survey reaches some 1.5\,mag. deeper than the sample of \citet{Wang2019}, the only other complete sample at these redshifts. Previous searches of VIKING \citep{Venemans2013,Venemans2015b} covered a smaller area, were not as deep, and were incomplete in that they did not follow up all candidates. 

We also undertook a comparison of three different selection methods, BMC, SED fitting and colour cuts. We found that the BMC method is the best. It reaches 0.3\,mag. deeper than the colour cuts method, while the number of false positives is a factor of eight smaller. The BMC method reaches 0.7\,mag. deeper than the SED-fitting method, which only finds one of the four known quasars in this field.

We find evidence for a population of emission line galaxies with strong [OIII]$\lambda\lambda$4959,5007 emission that are brighter examples of similar sources found in deeper quasar surveys near $z=6$. Such objects could potentially contaminate future surveys for high-redshift quasars with \textit{Euclid}, and therefore need to be better characterised.

In a forthcoming publication we will use the new complete sample and associated selection function to refine the estimated rate of decline in quasar space density over the interval $6<z<7.5$.

\section*{Acknowledgements}

Based on observations collected at the European Organisation for Astronomical Research in the Southern Hemisphere under ESO programme(s) 0102.A-0848(A). This publication makes use of data products from the Wide-field Infrared Survey Explorer, which is a joint project of the University of California, Los Angeles, and the Jet Propulsion Laboratory/California Institute of Technology, funded by the National Aeronautics and Space Administration. This publication also makes use of data products from NEOWISE, which is a project of the Jet Propulsion Laboratory/California Institute of Technology, funded by the Planetary Science Division of the National Aeronautics and Space Administration. We are very grateful to Christian Hummel our ESO support astronomer for extensive help in the preparation of the FORS2 spectroscopic observations. We thank Sophie Reed for useful discussions, and we thank the anonymous referee for comments which led to improvements in the clarity of the presentation.
This work was supported by grant ST/N000838/1 
from the Science and Technology Facilities Council. 
This research has benefitted from the SpeX Prism Spectral Libraries, 
maintained by Adam Burgasser at http://pono.ucsd.edu/~adam/browndwarfs/spexprism.
RB would like to thank L. Barnett-Kramp for useful comments on an early draft of the paper.

\section*{Data Availability}

This work is based on the VIKING data release VIKINGv20160406 available as part of the VISTA Science Archive (\url{http://horus.roe.ac.uk/vsa/index.html}).
A version of the BMC code is publicly available at \url{http://github.com/rhysrb/Pq_server}.
Raw ESO follow-up data can be accessed via \url{http://archive.eso.org/cms.html} as part of programme 0102.A-0848(A).
Derived data products generated in this research will be shared on reasonable request to
the corresponding author.



\bibliographystyle{mnras}
\bibliography{viking.bib} 




\appendix

\section{Primary and secondary observations in VIKING} 
\label{sec:PoS}

The basic VISTA survey area is, as described by \citet{Cross2012}, a `tile' (which we have also referred to as a frameset in this work),
formed of six individual exposures labelled `pawprints'.
These pawprints are offset such that a single tile is at least doubly exposed,
with the exception of a narrow strip at the top and bottom of the tile, which is imaged just once.
These strips overlap with adjacent tiles,
allowing the minimum survey depth to be achieved 
by ensuring the full VIKING area is imaged at least twice.
A consequence of this observation strategy is that duplicate entries appear in the table.
After source merging is complete, a process known as seaming takes place,
which identifies and flags duplicates, using the \texttt{PriOrSec} (POS) property.

Where a VIKING source is duplicated, the `better' entry 
as determined on the basis of the proximity to the optical axis of the camera,
is labelled as the primary using the POS flag, and the other entry labelled the secondary.
Our list-driven catalogue contains $4.3\times10^7$ entries,
of which 18 per cent are labelled as secondary.
This proportion corresponds closely to the area of a tile that is singly imaged \citep{Cross2012}.

Given that such a large proportion of VIKING sources are duplicated,
in our search for HZQs using the BMC method we applied slightly different criteria to these doubly-observed sources
to avoid missing any interesting candidates.
If either of the primary or secondary observations of a doubly-observed source satisfied $\pq > 0.1$,
we averaged the primary and secondary photometry using inverse variance weighting, 
and assessed its selection again, this time applying the $\pq > 0.15$ threshold.
Two known quasars (for which both the primary and secondary entries satisfied $\pq>0.15$ anyway) 
and two of the new candidates, indicated in Table~\ref{tab:bmccands}, were selected in this way.

We also assessed the impact of primaries/secondaries on our BMC HZQ selection function.
The flux errors of the VIKING secondaries are typically somewhat worse than the 
primary flux errors, up to a level of 30 per cent.
For each HZQ that we simulate (representing primaries),
we therefore produce a second set of photometry drawn from flux distributions with the uncertainties enhanced by 30 per cent (secondaries).
Again, if either the primary or secondary HZQ entry was selected using the relaxed criteria,
we produced a third set of photometry for each HZQ,
by averaging the primary and secondary photometry using inverse variance weighting.
We find, for a simulated HZQ, selection using the combined photometry 
matches selection using only primary data very closely.
We conclude that the presence of a secondary does not typically allow the selection of a source that is not otherwise selected,
nor does it prevent the selection of a HZQ that is selected on the basis of its primary.

\section{Population modelling}
\label{sec:pops}

Here we summarise the surface density terms and model colours 
which are used in the selection methods described in Sect.~\ref{sec:selection}.
We present the models for quasars in Sect.~\ref{sec:mq},
for MLTs in Sect.~\ref{sec:ms}, and for ETGs in Sect.~\ref{sec:mg}.
The population models are determined following the same procedures 
detailed in \citet{Barnett2019}, allowing for the differences between the $Euclid$ and VIKING filters, and the different image quality, which is relevant to modelling the ETGs.
In order not to repeat too much material between the two papers, the presentation is briefer in the current paper,
but we draw attention to aspects of the population models that differ from the $Euclid$ work.
The model colours are shown in Fig.~\ref{fig:colourcolour}.
As a reminder, in this paper magnitudes are quoted on the Vega system, whereas in \citet{Barnett2019} they are on the AB system.

\subsection{Quasars}
\label{sec:mq}

The parameters $\bm{\theta}$ for the quasars are absolute magnitude and redshift.
For the VIKING analysis we adopt the single power law QLF used by \citet{Mortlock2012}, 
since we do not probe significantly fainter than the ``knee" of the luminosity function \citep{Jiang2016,Matsuoka2018LF}.
The redshift evolution of this QLF uses $k=-0.47$:
we have not adopted the stronger values
since we started selecting VIKING quasar candidates 
before \citet{Jiang2016} or \citet{Wang2019} were published.
While using a weaker evolution for the quasar weight will boost the resulting values of \pq,
the relative probabilistic ranking of VIKING sources will be almost unchanged;
even using a more strongly evolving QLF we could select almost exactly the same list of candidates
by adjusting the \pq~threshold.

The nine quasar SEDs comprise combinations of three different continuum slopes and three different emission line strengths, and the details are provided in Section 5.6 of \citet{Barnett2019}.
As was the case in \citet{Barnett2019}, 
we use \texttt{synphot} to directly measure the quasar $k$-corrections and colours used in the SED fitting and BMC techniques,
from updated versions of the model spectra from \citet{Hewett2006} and \citet{Maddox2008}.
We use the version of the model with `standard' continuum slope and emission line strength in the selection.
For our simulated sources we assume that all flux blueward of \lya is absorbed, 
except that we include a near zone of radius 3\,Mpc (proper).
Our quasar selection functions are not sensitive to the exact choice of the near-zone size within the range $1-5$\,Mpc.

\begin{table}
	\centering
	\advance\leftskip-3cm
	\advance\rightskip-3cm
	\caption{MLT density and model colour data for VIKING.}
	\begin{adjustbox}{width=0.48\textwidth,totalheight=\textheight,keepaspectratio}
		\label{tab:mlt}
		\begin{tabular}{ccccccc}
			\hline \hline \\[-2ex]
			SpT & $\rho_0~(\mathrm{pc^{-3}})$ & $M_J$ & $Z-Y$ & $Y-J$ & $J-H$ & $H-\ks$\\ 
			\hline \\[-2ex]
			M0 & $2.4\times10^{-3}$ & 5.69 & 0.29 & 0.45 & 0.62 & 0.19\\
			M1 & $2.7\times10^{-3}$ & 6.24 & 0.32 & 0.47 & 0.60 & 0.22\\
			M2 & $4.4\times10^{-3}$ & 6.85 & 0.35 & 0.49 & 0.58 & 0.23\\
			M3 & $7.8\times10^{-3}$ & 7.38 & 0.39 & 0.51 & 0.55 & 0.24\\
			M4 & $1.0\times10^{-2}$ & 7.96 & 0.44 & 0.53 & 0.54 & 0.26\\
			M5 & $1.1\times10^{-2}$ & 8.56 & 0.51 & 0.57 & 0.53 & 0.29\\
			M6 & $7.8\times10^{-3}$ & 9.86 & 0.60 & 0.63 & 0.53 & 0.31\\
			M7 & $2.2\times10^{-3}$ & 10.65 & 0.70 & 0.68 & 0.54 & 0.34\\
			M8 & $1.7\times10^{-3}$ & 11.06 & 0.83 & 0.79 & 0.56 & 0.38\\
			M9 & $1.1\times10^{-3}$ & 11.31 & 0.99 & 0.87 & 0.59 & 0.42\\
			L0 & $6.7\times10^{-4}$ & 11.52 & 1.20 & 1.04 & 0.63 & 0.53\\
			L1 & $4.3\times10^{-4}$ & 11.78 & 1.29 & 1.11 & 0.67 & 0.55\\
			L2 & $3.8\times10^{-4}$ & 12.11 & 1.33 & 1.18 & 0.73 & 0.59\\
			L3 & $3.6\times10^{-4}$ & 12.51 & 1.36 & 1.23 & 0.79 & 0.62\\
			L4 & $5.3\times10^{-4}$ & 12.95 & 1.38 & 1.27 & 0.86 & 0.66\\
			L5 & $4.1\times10^{-4}$ & 13.40 & 1.40 & 1.31 & 0.91 & 0.68\\
			L6 & $2.2\times10^{-4}$ & 13.82 & 1.42 & 1.33 & 0.96 & 0.70\\
			L7 & $6.3\times10^{-4}$ & 14.17 & 1.46 & 1.35 & 0.97 & 0.70\\
			L8 & $3.9\times10^{-4}$ & 14.42 & 1.51 & 1.21 & 0.96 & 0.68\\
			L9 & $4.8\times10^{-4}$ & 14.56 & 1.59 & 1.20 & 0.90 & 0.63\\
			T0 & $6.3\times10^{-4}$ & 14.60 & 1.69 & 1.19 & 0.80 & 0.55\\
			T1 & $6.4\times10^{-4}$ & 14.55 & 1.82 & 1.19 & 0.65 & 0.45\\
			T2 & $3.6\times10^{-4}$ & 14.45 & 1.97 & 1.18 & 0.46 & 0.33\\
			T3 & $3.6\times10^{-4}$ & 14.35 & 2.14 & 1.18 & 0.25 & 0.19\\
			T4 & $5.6\times10^{-4}$ & 14.32 & 2.32 & 1.17 & 0.02 & 0.06\\
			T5 & $7.1\times10^{-4}$ & 14.44 & 2.52 & 1.16 & $-0.19$ & $-0.05$\\
			T6 & $2.1\times10^{-4}$ & 14.78 & 2.72 & 1.16 & $-0.35$ & $-0.12$\\
			T7 & $2.1\times10^{-3}$ & 15.42 & 2.92 & 1.15 & $-0.43$ & $-0.11$\\
			T8 & $7.5\times10^{-4}$ & 16.42 & 3.10 & 1.15 & $-0.36$ & 0.01\\
			\hline
		\end{tabular}
	\end{adjustbox}
\end{table}

\subsection{MLT dwarfs}
\label{sec:ms}

The total MLT weight is determined by summing individual sub-weights, computed for each spectral type \myrange{M0}{T8}.
We therefore require the full range of VIKING $ZYJH\ks$ colours for each spectral type.
\citet{Skrzypek2015,Skrzypek2016} provide UKIDSS $YJHK$ colours for types \myrange{M7}{T8}. 
The UKIDSS $YJH$ filters are a close match to VISTA;
however, there is no match in those works for the VISTA $Z$ and $\ks$ bands,
so \zy and \hk~must be determined separately for \myrange{M7}{T8}.
We additionally require the full set of VIKING colours for \myrange{M0}{M6}. 
To proceed, we select a bright subsample (${\rm S/N}_i$ > 50) from the spectroscopic M dwarf catalogue presented by \citet{West2011}. 
We determine average $Z - Y$ and $H - \ks$ colours for all M dwarfs by matching the M dwarf sample to VIKING.
We additionally match the M dwarf sample to UKIDSS and measure median $rizYJH$ SDSS/UKIDSS colours for \myrange{M0}{M5}.
These are required for the \myrange{M0}{M5} absolute magnitudes, as detailed further below.

For the L and T dwarf $\hk$ colour we make use of the corrections provided by \citet{Stephens2004}, 
which allow conversions between the MKO filter set used by UKIDSS, and other photometric systems. 
We find the DENIS $K$ band provides a close match to the VIKING $\ks$ filter, 
allowing us to approximate $\hk$ for L and T types. 
Finally, we determine $\zy$ colours for L and T dwarfs by fitting a polynomial to the sources presented by \citet{Hewett2006}.
We present the full set of VIKING $ZYJH\ks$ colours in Table~\ref{tab:mlt}.

The dwarf star surface density term requires number densities and absolute magnitudes for each spectral type.
We use the the Galactic plane number densities presented in \citet{Barnett2019}.
\citet{Dupuy2012} provide MKO $J$-band absolute magnitudes for spectral types \myrange{M6}{T8}. 
For \myrange{M0}{M5} we use the relationship between $i-z$ and $M_r$ from \citet{Bochanski2010}. 
We correct to the Vega system using the offsets provided by \citet{Hewett2006},
and then use the colours determined above to convert $M_r$ to $M_J$.
As in \citet{Barnett2019}, we have assumed MLT number density varies as $\rho=\rho_0\,e^{-Z/Z_s}$, 
where $\rho_0$ is the number density of any spectral type \myrange{M0}{T8} at the Galactic central plane,
$Z$ is the vertical distance from the plane, and $Z_s$ is the scale height, assumed to be 300\,pc. The small offset of the Sun from the Galactic plane
is disregarded.

\subsection{Early-type galaxies}
\label{sec:mg}

\begin{figure}
	\centering
	\includegraphics[width=9cm]{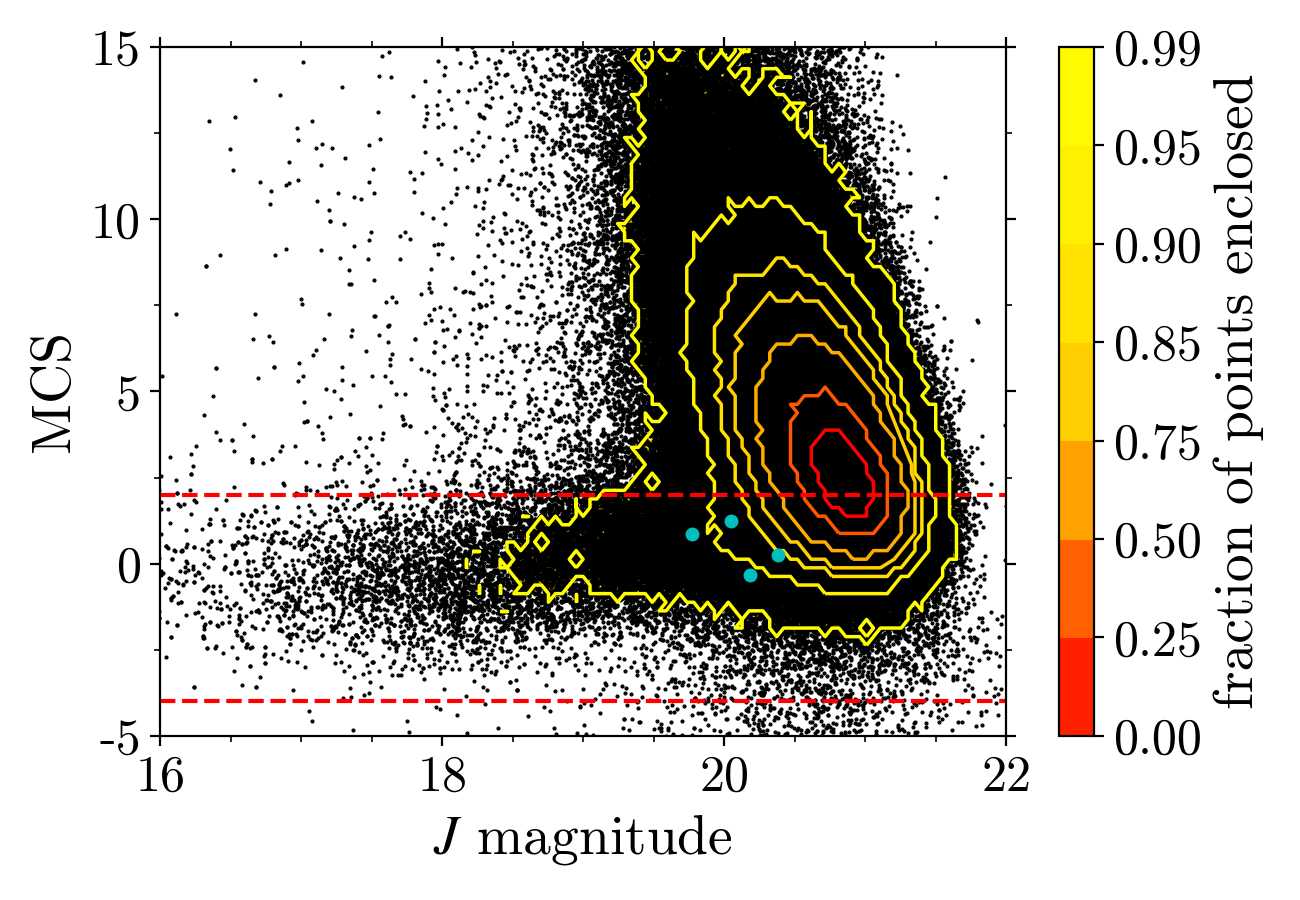}
	\caption{MCS statistic as a function of $J$ magnitude.
		Black points mark the sources in our VIKING sample, 
		where we have not taken a cut on MCS. The blue circles are the four HZQs.
		The red dashed lines indicate our cut to eliminate bright ellipticals.}
	\label{fig:mcs}
\end{figure}

Over the redshift range $z=\myrange{1}{2}$, ETGs have red NIR colours, 
which can resemble the colours of high-redshift quasars at low S/N.
In previous works, which have focused on the brightest candidates,
this source of contamination has been mitigated 
by taking a cut on a morphological statistic \citep[e.g.,][]{Mortlock2012,Venemans2013}.
However, size and stellar mass for the ETG population are strongly correlated \citep{Wel2014}, 
suggesting faint early-type galaxies at these redshifts will be very compact,
and classified as point sources by VIKING.
We illustrate this point in Fig.~\ref{fig:mcs},
which shows the behaviour of the MCS statistic as a function of $J$ magnitude for our initial sample
of red sources, before applying the cut on MCS (see Sect~\ref{sec:methods}).
At bright magnitudes ($J\lesssim19$), there are two distinct populations:
point sources, with $\mathrm{MCS}\simeq0$,
and extended sources at much higher values.
However, at fainter magnitudes, the two populations become merged,
i.e., morphology is no longer a good discriminator.
We therefore have to incorporate the ETG population as a contaminant.

The MCS statistic measures the similarity of the radial surface brightness profile to the profile of stars, and is normalised such that at any magnitude the distribution for point sources is Gaussian with standard deviation unity. As can be seen, at faint magnitudes the Gaussian distribution becomes more difficult to discern, and so to normalise correctly. However this is successfully achieved by fitting to the negative part of the distribution. The fact that for each of the four quasars MCS lies within the range $-2$ to $+2$ is an indication that the normalisation works well in the regime of high galaxy surface density.

We firstly derive a model for the surface density and colours of ETGs as a function of redshift and $J$ magnitude. We then account for the magnitude dependence of the MCS parameter.
As in \citet{Barnett2019}, we make use of the COSMOS/UltraVISTA sources presented by \citet{Laigle2016} to derive the model. 
\citet{Barnett2019} found that quiescent objects in the COSMOS catalogue
have a large range of formation redshifts ($z_f$),
and approximated the catalogue as two separate populations,
with a fraction 0.8 having $z_f = 3$, and a fraction 0.2 having $z_f = 10$.
We use colours computed for both formation redshifts,
from the evolutionary models of \citet[][]{Bruzual2003}. 

To produce a surface density model that can be applied to VIKING, we first fit a surface density model to  \ang{;;2.0} aperture COSMOS data. 
We then require corrections to convert COSMOS aperture magnitudes to VIKING APERMAG3 magnitudes.
The COSMOS/UltraVISTA area does not overlap with VIKING; 
however, there is overlap with the UKIDSS LAS, and the UKIDSS and VIKING data are similar in terms of image quality and the data-processing pipeline.
We find 195 COSMOS sources are detected in the $K$ band in the LAS,
allowing us to compare the \ang{;;2.0} aperture magnitudes between the LAS and COSMOS.
The UKIDSS APERMAG3 photometry corresponds to the flux in a \ang{;;2.0} aperture, that is then aperture corrected to total flux using the aperture correction for a point source. Therefore the aperture correction has to be subtracted to get the LAS flux in the \ang{;;2.0} aperture. After doing this we find
the UKIDSS fluxes agree with the COSMOS fluxes almost exactly, on average. On this basis we use the following relation between the \ang{;;2.0} aperture $J$-band AB COSMOS magnitudes $J_\mathrm{corr}$
and VIKING Vega APERMAG3 $J$ magnitudes:
$J_\mathrm{corr} = J + 1.147$. This equation incorporates both
the AB correction and the point source aperture correction.

The surface density function 
in terms of $J_\mathrm{corr}$ and source redshift is determined from a maximum likelihood fit.
The functional form of this function 
in units of $\mathrm{mag}^{-1}\,\mathrm{deg}^{-2}$ per unit redshift is
\begin{equation}
\Sigma(J_\mathrm{corr},z) = 
\alpha\,
\exp\left\lbrace-\frac{1}{2}\left[\frac{J_\mathrm{corr} - \left(J_0+b\,z\right)} {\sigma}\right]^2\right\rbrace\,
\exp\left[-\left(\frac{z-0.8}{z_0}\right)\right]\\
\label{eq:gpr}
\end{equation}
where we find the best-fitting parameters to be 
$\left(\alpha,\sigma,J_0,b,z_0\right) = \left(7697,0.883,20.467,1.462,0.429\right)$.
We assume the same function is applicable to early-type galaxies with either formation redshift,
and scale the resulting weights by 0.8 for $z_f=3$ and 0.2 for $z_f=10$
to reflect the distribution of $z_f$ values seen in the COSMOS data.

\begin{figure}
	\centering
	\includegraphics[width=9cm]{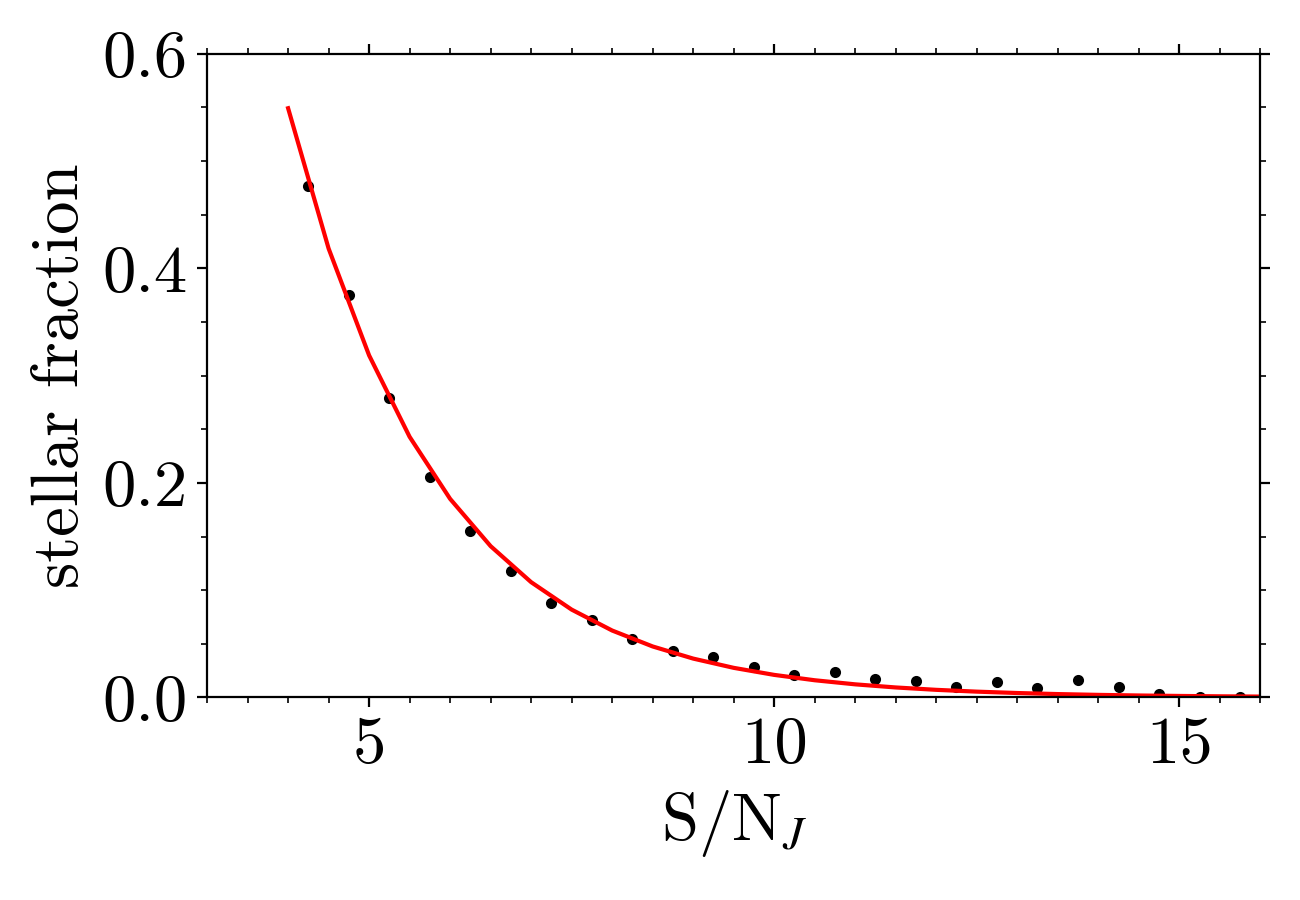}
	\caption{Measured fraction of ETGs classified as stellar as a
		function of $\sn_J$. The red curve shows the exponential fit, Eq.~\ref{eq:mcs}.}
	\label{fig:morphsn}
\end{figure}

The surface density function and the model colours provide a complete description of the ETG population, with the exception that 
only objects satisfying the cut $-4<\mathrm{MCS}<2$ will appear in the sample. We therefore need to quantify the magnitude dependence of this cut on MCS.
We wish to evaluate the fraction of ETGs that satisfy the MCS cut. 
From the counts of red sources as a function of \sn we can tally separately stellar sources, with $-4<\mathrm{MCS}<2$, and extended sources. We can subtract the
counts of MLTs from the stellar counts using the density model for these sources. Then we are left with galaxy counts only as a function of \sn, separated into galaxies classified as stellar and galaxies classified as extended. The proportion of galaxies classified as stellar as a function of \sn in the $J$ band is plotted in Fig.~\ref{fig:morphsn}. 
The resulting function is well fit by an exponential of the form 
\begin{equation}
\mathrm{stellar~fraction} = a\cdot\exp\left(-b\cdot\sn_J\right),
\label{eq:mcs}
\end{equation}
where we find the best fit to be $(a,b) = (4.837,0.544)$.
The full galaxy prior is formed by multiplying Equations~\ref{eq:gpr} and~\ref{eq:mcs}.

\section{Simulating detection in the $J$ band}
\label{sec:casu}

\begin{figure}
	\centering
	\includegraphics[width=9cm]{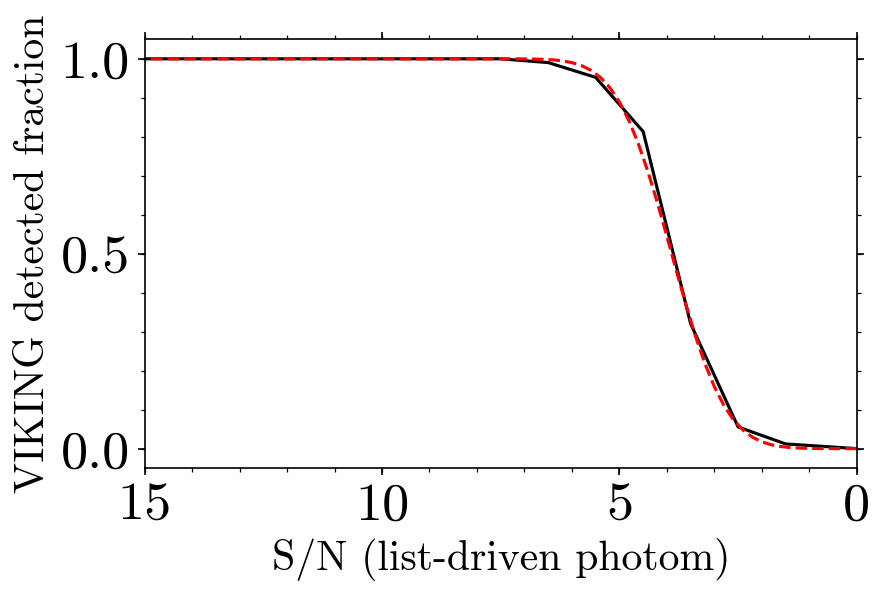}
	\caption{VIKING completeness as a function of S/N in any band. 
		Results from the $Y,H$ and $\ks$ bands have been averaged. 
		The red dashed line plots the cumulative Gaussian function that we fit to the data.}
	\label{fig:video}
\end{figure}

Our list-driven sample starts with the list of objects detected in the original VIKING catalogue which have $\mathrm{S/N}>4$ in an aperture of diameter \ang{;;2.0}. For a source in the $J$ band that has a given S/N in the aperture, we want to know the probability  that it would have been detected, i.e., would have made it into the VIKING catalogue. We cannot measure this in the $J$ band itself since we do not have measurements of the sources not in the VIKING catalogue. 
However, whether or not a source in, say, the $Y$ band measured at a particular S/N gets into the VIKING catalogue does not depend on what happens in the $J$ band. So we can gauge the detection probability in any chosen band starting with any catalogue of real sources detected in some other band, and then measuring them in the chosen band. 

We start by matching a sample of sources classified as stellar in the deeper 
VISTA Deep Extragalactic Observations survey \citep[VIDEO;][]{Jarvis2013} survey to our list-driven catalogue. This simply ensures that the objects are real, since they are found in both catalogues. Then in, e.g., the $Y$ band, we evaluate the fraction of sources that are detected in $Y$ in VIKING (i.e., appear in the original VIKING catalogue) as a function of S/N in the aperture in that band. The results averaged over the $Y$, $H$, and $\ks$ bands are plotted in Fig.~\ref{fig:video}. 

To be clear, this says that if an object has the given S/N in the aperture on any frame, then the probability that it will appear in the original VIKING catalogue is given by the function plotted. This is the function needed to quantify detection in the $J$ band for modelling the selection functions (Sect.~\ref{sec:simulations}), and for creating the synthetic catalogue (Sect.~\ref{sec:mltell}). We fit a Gaussian cumulative distribution function (CDF; i.e., the error function) to this curve. The (50,68,95) per cent quantiles are found to lie at at $\sn = (3.9,4.8,5.7)$. 
Each source that we simulate is then detected or rejected with a probability set by this CDF,
leaving us with sources that would be present in the VIKING list-driven catalogue.

\bsp	
\label{lastpage}
\end{document}